\documentclass[aps,pra,twocolumn,superscriptaddress,nofootinbib]{revtex4-1}

\usepackage{lipsum}
\usepackage{braket}
\usepackage{graphicx}% Include figure files
\usepackage{dcolumn}% Align table columns on decimal point
\usepackage{bm}% bold math
\usepackage{amsmath}
\usepackage{subfigure}
\usepackage{makecell}
\usepackage{verbatim}
\usepackage{color}

\newcommand{\abs}[1]{\lvert #1 \rvert}
\newcommand{\ev}[1]{\langle #1 \rangle}
\newcommand{\lv}[0]{\mathcal{L}}
\newcommand{\mP}[0]{\mathcal{P}}%for partial sums
\newcommand{\mQ}[0]{\mathcal{Q}}%for ion partial sums

\newcommand{\nbar}[0]{\bar{n}}

\newcommand{\mbf}[1]{\mathbf{#1}}

\newcommand{\COM}[0]{\text{COM}}

\begin{document}

\title{Modeling near ground-state cooling of two-dimensional ion crystals in a Penning trap using electromagnetically induced transparency}

\author{Athreya Shankar}
\email[]{athreya.shankar@colorado.edu}
\affiliation{JILA, NIST, and Department of Physics, University of Colorado Boulder, Boulder, Colorado 80309, USA}

\author{Elena Jordan}
\affiliation{Time and Frequency Division, National Institute of Standards and Technology, Boulder, Colorado 80305, USA}

\author{Kevin A. Gilmore}
\affiliation{JILA, NIST, and Department of Physics, University of Colorado Boulder, Boulder, Colorado 80309, USA}
\affiliation{Time and Frequency Division, National Institute of Standards and Technology, Boulder, Colorado 80305, USA}

\author{Arghavan Safavi-Naini}
\affiliation{JILA, NIST, and Department of Physics, University of Colorado Boulder, Boulder, Colorado 80309, USA}

\author{John J. Bollinger}
\affiliation{Time and Frequency Division, National Institute of Standards and Technology, Boulder, Colorado 80305, USA}

\author{Murray J. Holland}
\affiliation{JILA, NIST, and Department of Physics, University of Colorado Boulder, Boulder, Colorado 80309, USA}

\date{\today}

\begin{abstract}

Penning traps, with their ability to control planar crystals of tens to hundreds of ions, are versatile quantum simulators. Thermal occupations of the motional drumhead modes, transverse to the plane of the ion crystal, degrade the quality of quantum simulations. Laser cooling using electromagnetically induced transparency (EIT cooling) is attractive as an efficient way to quickly initialize the drumhead modes to near ground-state occupations. We numerically investigate the efficiency of EIT cooling of planar ion crystals in a Penning trap, accounting for complications arising from the nature of the trap and from the simultaneous cooling of multiple ions. We show that, in spite of challenges, the large bandwidth of drumhead modes (hundreds of kilohertz) can be rapidly cooled to near ground-state occupations within a few hundred microseconds. Our predictions for the center-of-mass mode include a cooling time constant of tens of microseconds and an enhancement of the cooling rate with increasing number of ions. Successful experimental demonstrations of EIT cooling in the NIST Penning trap [E. Jordan, K. A. Gilmore, A. Shankar, A. Safavi-Naini, M. J. Holland, and J. J. Bollinger, ``Near ground-state cooling of two-dimensional trapped-ion crystals with more than 100 ions", (2018), arXiv:1809.06346] validate our predictions.

\end{abstract}

%\keywords{}

\maketitle

\section{\label{sec:introduction}Introduction}

Trapped ions have rapidly evolved to become a leading platform for quantum computing, quantum simulation and metrology \cite{haffner2008PhysRep,blatt2012Nat}. Specifically, ions stored in Penning traps have been demonstrated to be 
ideal for analog quantum simulation as well as quantum-enhanced sensing, in part because large ion crystals are routinely formed and controlled in this device \cite{biercuk2009QIC, mavadia2013Nat, ball2018arxiv}. For example, planar crystals 
of tens to hundreds of ${}^9\text{Be}^+$ ions have been used to simulate spin-spin as well as spin-boson models including the Ising \cite{britton2012Nat,bohnet2016Sci}, transverse-field Ising, as well as Dicke models \cite{safavi2018PRL,cohn2018NJP}. Quantum information studies on the growth of entanglement \cite{garttner2017Nat} as well as investigations on preparing ground states of exotic Hamiltonians \cite{safavi2018PRL,cohn2018NJP} have shown that exciting many-body physics can be studied with this versatile quantum simulator. In addition, ions in Penning traps serve as excellent motion sensors capable of resolving, in a single experimental trial, motional amplitudes smaller than the zero-point fluctuations of the normal modes dictating the motion transverse to the crystal plane \cite{gilmore2017PRL}, thus enabling the detection of extremely weak forces and electric fields. 

For implementing these protocols with the NIST Penning trap, the spin is encoded in two hyperfine ground states of ${}^9\text{Be}^+$ \cite{biercuk2009QIC}. Spin-spin interactions are mediated by the motional drumhead modes, transverse to the crystal plane, that arise from the interplay of the trap potential and the inter-ion Coulomb repulsion, with the spin-motion coupling generated using suitable drive lasers \cite{britton2012Nat}. As a result, excess thermal energy in these normal modes adversely affects the science protocol being investigated, higlighting the need for sub-Doppler, near ground-state cooling. For example, the fidelity of preparing the ground state of the Dicke model is significantly reduced by the thermal occupation of $\nbar \approx 6$ of the center-of-mass (COM) mode, which is close to the Doppler cooling limit \cite{safavi2018PRL,cohn2018NJP}. Estimates show that the fidelity significantly improves if the COM mode is cooled down to $\nbar \approx 0$. Further, near ground-state cooling  should also greatly improve the motion sensing capability of this platform. 

\begin{figure}[!htb]
\centering
\includegraphics[width=\linewidth]
{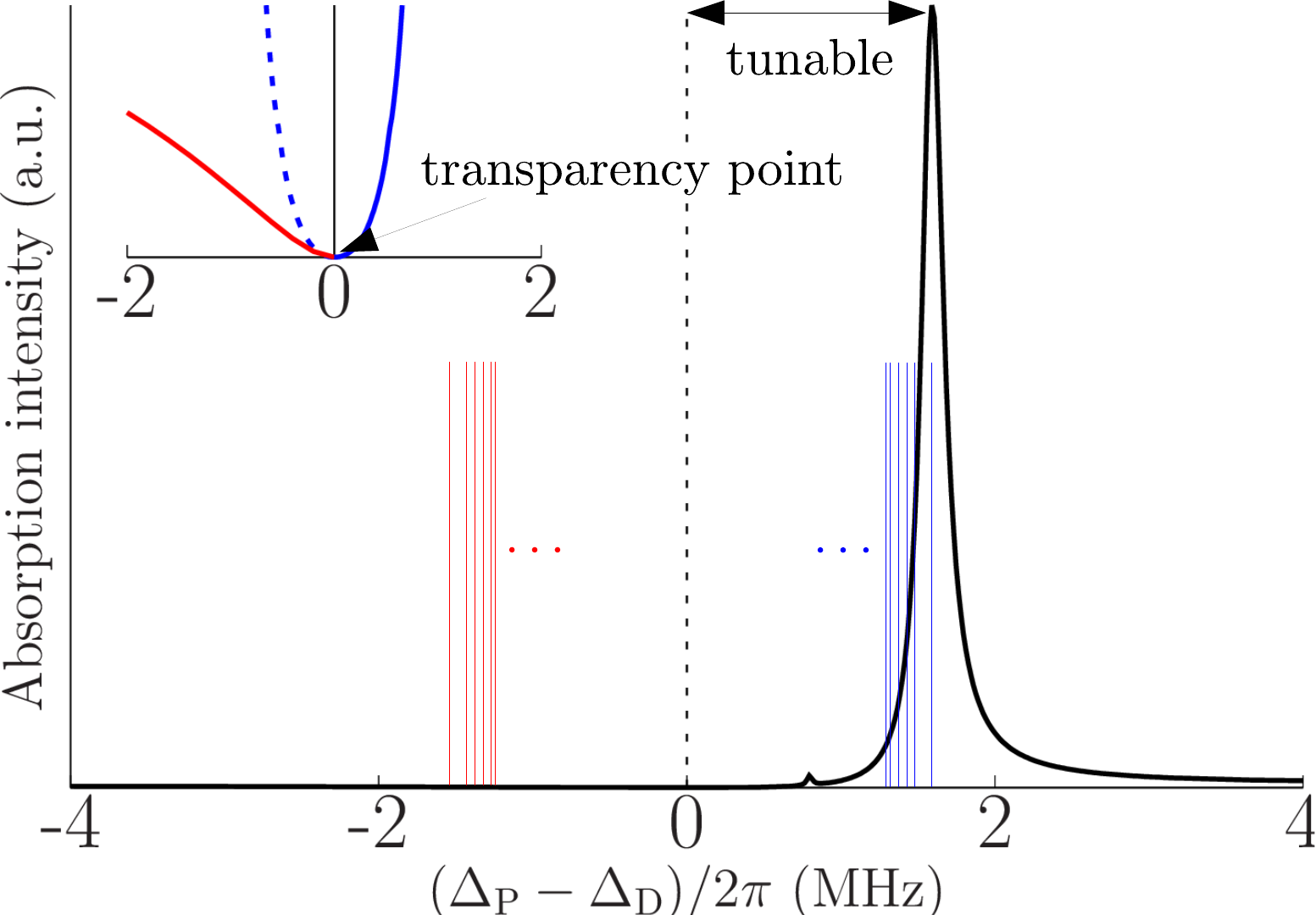}
\caption{(color online) EIT absorption spectrum with laser parameters relevant to the NIST EIT cooling experiment. Two strong dressing lasers, with equal Rabi frequencies and equal detunings from their respective transitions, couple two long-lived states to an excited state in a closed three-level system. The absorption from a weak probe is plotted as the probe detuning $\Delta_\text{P}$ is scanned across the dressing detuning $\Delta_\text{D}$. The motion-removing (blue impulses) and motion-adding (red impulses) sidebands can be interpreted as weak probes that sample this spectrum. Inset: Close-up near $(\Delta_\text{P}-\Delta_\text{D})/2\pi=0$, with a magnified $y$-axis, showing the zero at the transparency point as well as the asymmetry in the spectrum on either side of this point. The blue dashed line is a mirror image of the spectrum on the blue-detuned side, drawn on the red-detuned side to highlight the asymmetric growth of the absorption away from the transparency point. The dressing lasers have equal detuning $\Delta_\text{D}/2\pi \equiv \Delta^0/2\pi = 360$ MHz, and equal Rabi frequency $\Omega_\text{opt}(\Delta^0)/2\pi \approx 33.9$ MHz. The Rabi frequency of the weak probe is $\Omega_\text{P} = 0.05 \; \Omega_\text{opt}(\Delta^0)$. The decay rates from the excited state to the two long-lived states are $\Gamma_1/2\pi = 6$ MHz, $\Gamma_2/2\pi = 12$ MHz. (See the discussion in Section~\ref{sec:setup} and Eq.~(\ref{eqn:opt_rabi}) for a detailed explanation of the parameters.)}
\label{fig:eit_fano}
\end{figure}

Electromagnetically induced transparency (EIT) promises a path for cooling the 
entire bandwidth of drumhead modes close to their ground states. In contrast to sideband cooling where the modes are cooled one-by-one by sweeping the two-photon detuning across the bandwidth of modes, EIT cooling can potentially cool the full bandwidth of modes in a single experimental application with no time-varying parameters, allowing for simpler implementation and faster cooling. The naive expectation comes from the well understood physics of EIT cooling of a single trapped ion \cite{morigi2003PRA}, which we now recall briefly. The ion is assumed to have a closed three-level electronic manifold consisting of two long-lived states, such as the hyperfine ground states of ${}^9\text{Be}^+$, and an excited state (see Fig.~\ref{fig:eit_expt_setup}(b)). Two strong dressing lasers couple the long-lived states to the excited state and are equally blue detuned from their respective transitions. EIT cooling can be understood by considering the absorption of a fictitious weak probe coupling one of the long-lived states to the excited state. As shown in Fig.~\ref{fig:eit_fano}, the steady-state absorption spectrum has a unique profile as the probe detuning $\Delta_\text{P}$ is swept, with the absorption exactly vanishing when the probe detuning equals the dressing detuning $\Delta_\text{D}$. A sharp peak immediately follows this transparency point and the separation between this peak and the transparency point can be tuned using the dressing laser powers. For a trapped ion, the motion-adding and motion-removing sidebands of the dressing lasers serve as weak probes that sample this absorption spectrum. Tuning the separation between the sharp peak and the transparency point to match the motional frequency causes the motion-removing sideband to be strongly enhanced and the motion-adding sideband to be strongly suppressed, leading to highly efficient cooling. In the Penning trap, by tuning this separation to coincide with the frequency of the COM mode, which is the highest frequency drumhead mode, all the drumhead modes have highly asymmetric motion-removing and motion-adding sidebands, which should produce efficient cooling over the full bandwidth of these modes. In this paper, we theoretically investigate this idea under realistic experimental conditions employed in the NIST Penning trap.   

The unique challenges confronted in implementing EIT cooling in a Penning trap further motivate our  theoretical study of the prospects for its success. First, ions stored in a Penning trap are constantly revolving around the trap center and therefore, in general, experience time-varying Doppler shifts on the applied dressing lasers. Second, experimental constraints as well as a compromise between the speed of cooling and the final temperature dictate that the timescales for the electronic and motional degrees of freedom may not be sufficiently well separated to adiabatically eliminate the electronic degrees of freedom, as was done in the initial analysis of trapped-ion EIT cooling \cite{morigi2003PRA,morigi2000PRL}. Third, although EIT cooling has been used to cool all the radial modes of a linear chain of up to 18 ions \cite{lechner2016PRA}, the dynamics of simultaneously EIT cooling several tens to hundreds of normal modes that can interact via the applied lasers is not well understood and could be very different from the single-ion case. 

At the outset, we summarize the major predictions from our study. Under typical experimental conditions, EIT cooling leads to near ground-state occupancies for all the drumhead modes, spread over a bandwidth of hundreds of kilohertz, of large ion crystals in a Penning trap. The cooling of the COM mode has a time constant of few tens of $\mu$s. Further, the cooling rate of the COM mode increases with the number of ions in the crystal. These predictions have been verified by the successful demonstration of EIT cooling with more than 100 ions \cite{jordan2018}, where significant sub-Doppler cooling, strongly suggestive of near-ground state cooling, has been observed over the full bandwidth of drumhead modes. Quantitative measurements on the COM mode reveal occupations of $\bar{n} \approx 0.3 \pm 0.2$, and a measured cooling constant $\tau \approx 28 \; \mu$s. The measured cooling rate is faster than the expected single-ion rate under the same experimental conditions.

This paper is organized as follows. In Section \ref{sec:setup}, we describe the NIST Penning trap and proceed to set up the master equation for the EIT cooling of multiple ions. In Section \ref{sec:single_ion}, we use a toy model of a single revolving ion to illustrate the degrading effects of the time-varying Doppler shifts as well as demonstrate the invalidity of the adiabatic elimination procedure in our system. In Section \ref{sec:multi_ion}, we first build a Gaussian model for approximately studying the cooling dynamics for multiple ions, and demonstrate the near-ground state cooling of the COM mode for crystals with up to $N = 37$ ions. By comparing the cooling transients from our Gaussian model to exact calculations for a single ion, we show the build-up of beyond Gaussian correlations between the system degrees of freedom, which we are able to reproduce by systematically extending our approximate model beyond the Gaussian regime. Our improved model predicts a surprising enhancement in the cooling rate of the COM mode with increasing number of ions that is not captured by the Gaussian model. In Section \ref{sec:full_bandwidth}, we show how EIT cooling works efficiently over the full bandwidth of drumhead modes of crystals with as many as 120 ions, resulting in near ground state occupancies for all these modes. In Section \ref{sec:misalign}, we briefly demonstrate the expected robustness of EIT cooling to small misalignments of the dressing lasers. We conclude with a brief summary in Section \ref{sec:conclusion}, where we also discuss possible future extensions of our work.

\section{\label{sec:setup} Modeling the experiment}

\subsection{NIST Penning trap}

The NIST Penning trap is used to routinely produce, control and manipulate planar ion crystals of tens to hundreds of ${}^9\text{Be}^+$ ions \cite{biercuk2009QIC}. A static electric quadrupole field is used to achieve transverse confinement, while the addition of a strong transverse magnetic field ensures radial confinement. Potentials applied to electrodes arranged symmetrically around the $z$-axis generate the required electric fields and a superconducting magnet produces a strong magnetic field of $4.46 \; \text{T}$. The $\mbf{E}\times\mbf{B}$ drift of the ions arising from the combination of the magnetic field and the azimuthally symmetric electric fields causes the ions to revolve around the trap center. The frequency of the crystal rotation can be precisely controlled and stabilized by applying a weak rotating potential on the electrodes. Typically, this `rotating wall' potential is used to stabilize the rotation frequency of the crystal to $\omega_r/2\pi = 180 \; \text{kHz}$. When the radial confinement is weak compared to the transverse confinement, the ions form a 2D planar crystal with an approximate triangular lattice 
(see Fig.~\ref{fig:eit_expt_setup} for an illustration). The strength of the transverse harmonic confinement is characterized by a trapping frequency that is also the frequency of the transverse center-of-mass (COM) mode, $\omega_\text{COM}$, of the ion crystal. A planar ion crystal with $N$ ions has $3N$ normal modes of motion, $2N$ of which are in-plane modes superposed on the crystal rotation, and $N$ of which are drumhead modes transverse to the crystal plane. The COM mode is the highest-frequency drumhead mode. For the experiments we model here, $\omega_\text{COM}/2\pi \approx 1.57 - 1.59 \; \text{MHz}$, tunable using trap parameters.  
\subsection{Master equation model}

\begin{figure}[!htb]
\centering
\includegraphics[width=0.8\linewidth]
{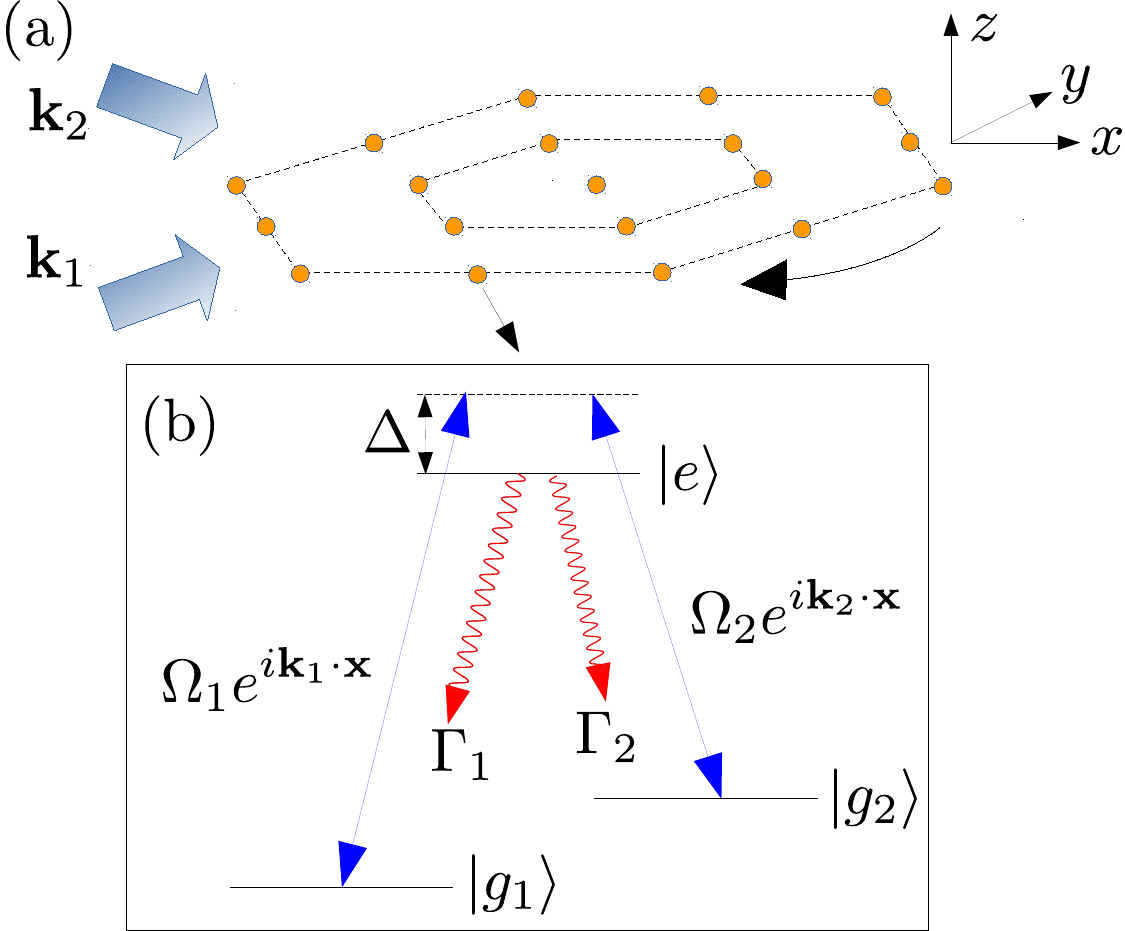}
\caption{(color online) Experimental setup to cool ions in a Penning trap using electromagnetically induced transparency (EIT). Two EIT lasers address the ion crystal at angles $\pm \theta$ with respect to the $x$-axis, with one driving the $\ket{g_1}\leftrightarrow\ket{e}$ transition and the other driving the $\ket{g_2}\leftrightarrow\ket{e}$ transition in a blue-detuned regime ($\Delta > 0$). The curved arrow indicates the rotation direction in the $x$-$y$ plane.}
\label{fig:eit_expt_setup}
\end{figure}

We consider singly-charged positive ions with a closed three-level electronic structure loaded in a Penning trap (see Fig.~\ref{fig:eit_expt_setup}). The two hyperfine ground states of each ion are labeled $\ket{g_1}$ and $\ket{g_2}$. Two EIT lasers are respectively blue-detuned from the $\ket{g_1} \leftrightarrow \ket{e}$ and $\ket{g_2}\leftrightarrow\ket{e}$ transitions, where $\ket{e}$ is an excited state separated from the two ground states by optical frequencies. The two EIT lasers are incident on the planar ion crystal at angles $\pm \theta$ with respect to the plane of the crystal, which we take to be the $x$-$y$ plane. The $\{\ket{e},\ket{g_1},\ket{g_2}\}$ manifold is a closed system, with decay rates of $\Gamma_1$ and $\Gamma_2$ for the $\ket{e}\rightarrow\ket{g_1}$ and $\ket{e}\rightarrow\ket{g_2}$ pathways respectively. We will adopt the shorthand notation $\sigma_{\alpha\beta}$ to denote the internal state operator $\ket{\alpha}\bra{\beta}$.

In the Schr{\"o}dinger picture, the Hamiltonian for the interaction of the EIT lasers with a single ion is $H^\text{S} = H_0^\text{S} + H_1^\text{S}(t)$, where  
\begin{equation}
H_0^\text{S} = -\sum_\mu \left(\omega_e-\omega_{g_\mu}\right) \sigma_{g_\mu g_\mu}
\end{equation}

\noindent and
\begin{eqnarray}
H_1^\text{S}(t) = &&\sum_\mu \frac{\Omega_{\mu}}{2} \left( e^{i \left( \mbf{k}_{\mu} \cdot \mbf{r}(t) - \omega_\mu t \right)} \sigma_{eg_\mu} + \text{H.c.} \right),
\end{eqnarray}

\noindent where the index $\mu=1,2$ accounts for the two ground states and the two EIT lasers coupling them to $\ket{e}$. Here, $\omega_e - \omega_{g_{\mu}}$ are the $\ket{g_{\mu}}\leftrightarrow\ket{e}$ transition frequencies, $\Omega_{\mu}$ are the Rabi frequencies for the laser-ion interaction, and $\omega_{\mu}$ and $\mbf{k}_{\mu}$ are the angular frequencies and propagation vectors for the two EIT lasers. Eventually, we will describe the motion along the $z$-direction using a set of quantized normal modes, and add the self-energy terms associated with these quantum harmonic oscillators to the interaction Hamiltonian. Throughout this paper, we have set $\hbar = 1$, unless we explicitly specify otherwise.   

Since any given ion in the Penning trap is undergoing a coherent rotation in the $x$-$y$ plane, its position $\mbf{r}(t)$ is 
\begin{eqnarray}
\mbf{r}(t) = &&\left(x(0) + \int_0^t dt' v_x(t') \right) \mbf{\hat{x}} \nonumber\\*
           + &&\left(y(0) + \int_0^t dt' v_y(t') \right) \mbf{\hat{y}} 
           + z(t)\mbf{\hat{z}}, 
\end{eqnarray}

Here we are neglecting any thermal motion in the plane of the ion crystal and assume $v_x(t')$ and $v_y(t')$ arise from the coherent circular motion caused by the $\mbf{E} \times \mbf{B}$ drift, which is the dominant in-plane motion. We assume that the laser beams are propagating in the x-z plane, so that $\mbf{k}_1 = k_{1,x}\mbf{\hat{x}} + k_{1,z}\mbf{\hat{z}}$ and $\mbf{k}_2 = k_{2,x}\mbf{\hat{x}} + k_{2,z}\mbf{\hat{z}}$. 

We then transform to an interaction picture with a time-dependent `free evolution' Hamiltonian $H_f^\text{S}(t)$ given by 
\begin{eqnarray}
H_f^\text{S}(t) = &&-\sum_\mu \left\{\omega_\mu - k_{\mu,x}v_x(t) \right\}\sigma_{g_\mu g_\mu}. 
\end{eqnarray}

\noindent The interaction picture transformation is very similar to the usual case with a time-independent free evolution Hamiltonian because $H_f^\text{S}(t)$, $H_f^\text{S}(t')$ commute at all times $t,t'$. The interaction Hamiltonian in this frame, given by the transformation $H^\text{I}(t) = e^{i\int_0^t dt' H_f^\text{S} (t')} (H_0^\text{S}+H_1^\text{S}(t) - H_f^\text{S}(t)) e^{-i\int_0^t dt' H_f^\text{S} (t')}$, is 
\begin{eqnarray}
H^\text{I}(t) && = \sum_\mu \Delta_\mu(t) \sigma_{g_\mu g_\mu} \nonumber\\*
         &&+\sum_\mu \frac{\Omega_\mu}{2}\left\{ e^{ik_{\mu,x}x(0)}e^{ik_{\mu,z}z(t)} \sigma_{eg_\mu} + \text{H.c.}\right\},
\label{eqn:int_Hamiltonian_1}
\end{eqnarray}

\noindent where $\Delta_{\mu}(t) = \Delta_{\mu}^0 - k_{\mu,x}v_x(t)$ are the effective detunings of the EIT lasers as seen by the ion, with $\Delta_{\mu}^0 = \omega_{\mu}-(\omega_e-\omega_{g_{\mu}})$. For an ion with initial position 
$\left\{x(0), y(0)\right\}$, the $x$-component of the velocity is $v_x(t) = \omega_r (y(0) \cos \omega_r t - x(0)\sin \omega_r t )$, where $\omega_r$ is the angular frequency of the rotating wall potential.

To perform EIT cooling, we tune the EIT lasers to satisfy the two-photon resonance on the blue-detuned side \cite{morigi2003PRA}, i.e. $\Delta_1^0 = \Delta_2^0 \equiv \Delta^0 > 0$. Further, the lasers are aligned such that their difference wavevector lies along the $z$-axis. This implies $k_{2,x}=k_{1,x}$ and $k_{2,z}\approx -k_{1,z}$. However, we will develop the theory without these two simplifications, and only apply these conditions numerically. For $N$ ions in the Penning trap, the interaction Hamiltonian generalizes straightforwardly as a sum over all ions.

The equilibrium crystal configuration results from the balance of the trap potential and the inter-ion Coulomb repulsion. The transverse motion of the $N$ ions about their equilibrium positions are not independent, instead being described by a set of $N$ collective normal modes with frequencies $\omega_n$, $n=1,2,\ldots, N$ and amplitudes $\mathcal{M}_{jn}$ at each ion $j$. The frequencies $\omega_n$ and the column vectors of the matrix $\mathcal{M}$ are respectively obtained as the eigenvalues and eigenvectors of the potential energy matrix associated with the coupled transverse harmonic motion of the ions \cite{wang2013PRA}. The transverse displacement of any ion $j$ can be expressed in terms of the $N$ quantized drumhead modes of the ion crystal as 
\begin{equation}
z_j(t) = \sum_{n=1}^N \sqrt{\frac{\hbar}{2M\omega_n}}\mathcal{M}_{jn} \left( b_n e^{-i\omega_n t} + b_n^\dag e^{i \omega_n t} \right),
\label{eqn:z_of_t}
\end{equation} 

\noindent where $b_n^\dag, b_n$ are the creation and annihilation operators for the normal mode $n$. 

The time-dependent exponentials in Eq. (\ref{eqn:z_of_t}) can be recast as self-energy terms, leading to the total interaction Hamiltonian 
\begin{eqnarray}
H^\text{I}(t) && = \sum_n \omega_n b_n^\dag b_n 
       + \sum_{j,\mu} \Delta_{\mu,j}(t) \sigma_{g_\mu g_\mu}^j  \nonumber\\*
&& + \sum_{j,\mu} \frac{1}{2} \left\{ \Omega_{\mu,j} e^{ik_{\mu,z}z_j} \sigma_{eg_\mu}^j + \text{H.c.}\right\},
\label{eqn:total_int_Hamiltonian}
\end{eqnarray}

\noindent where the displacement operators $z_j$ are now simply $
z_j = \sum_n \sqrt{\hbar/2M\omega_n}\mathcal{M}_{jn} \left( b_n + b_n^\dag \right)$.

Here, the instantaneous detunings experienced by each ion is different, depending on the $x$-component of the ion's velocity at that time point. In writing Eq.~(\ref{eqn:total_int_Hamiltonian}), the complex phase factors associated with the initial positions of the ions have been absorbed into the (now complex) Rabi frequencies. Further, for large ion crystals, the spatial profile of the EIT lasers over the extent of the crystal may be important, and therefore $\Omega_{1(2),j} \equiv \Omega^0_{1(2)} \left( x_j(t),y_j(t) \right) \times e^{ik_{1(2),x}x_j(0)}$, i.e. the amplitude of the Rabi frequency is in general a function of the instantaneous in-plane position of the rotating ion. 

Spontaneous emission from the excited level $\ket{e}$ to the two ground states is accounted for using dissipation terms written in Lindblad form, i.e. for any jump operator $O$, the dissipation term takes the form ${ \mathcal{D}[O]\rho = O \rho O^\dag - \frac{1}{2} O^\dag O \rho - \frac{1}{2} \rho O^\dag O }$, where $\rho$ is the density matrix of the system at hand. Since we are interested in the effect on the motion along the $z$-direction, the dissipation terms must account for the recoil momentum along the $z$-axis due to spontaneous emission \cite{dalibard1985JPhysB}. Therefore, the Lindblad term for ion $j$, including recoil associated with spontaneous decay of $\ket{e}$ to $\ket{g_\mu}$, is 
\begin{eqnarray}
\mathcal{D}_{\mu,j} \rho =  \Gamma_\mu && \left\{ \int_{-1}^{1} du \mathcal{N}_\mu (u) \sigma_{g_\mu e}^j e^{-i k_{\text{sc}} z_j u} \rho e^{i k_{\text{sc}} z_j u} \sigma_{eg_\mu}^j  \right. \nonumber\\* 
&&  \left. - \frac{1}{2}\sigma_{ee}^j\rho -\frac{1}{2}\rho\sigma_{ee}^j \right\},
\label{eqn:lindblad_d1}
\end{eqnarray}

\noindent where $\Gamma_\mu$ is the spontaneous decay rate of $\ket{e}\rightarrow\ket{g_\mu}$, $\mbf{k}_{\text{sc}}$ is the wavevector associated with the spontaneously emitted photon, $u = \cos \theta_\text{sc}$, with $\theta_\text{sc}$ the angle between $\mbf{k}_{\text{sc}}$ and the $z$-axis, and $\mathcal{N}_\mu(u)$ is the normalized dipole radiation pattern associated with the transition. Finally, the master equation for EIT cooling of $N$ ions in a Penning trap is given by 
\begin{equation}
\dot{\rho} = -i [H^\text{I}(t),\rho] + \sum_{j,\mu} D_{\mu,j} \rho, 
\label{eqn:master_eqn_full} 
\end{equation}

\noindent where $H^\text{I}(t)$ and $D_{\mu,j}\rho$ are as in Eq.~(\ref{eqn:total_int_Hamiltonian}) and Eq.~(\ref{eqn:lindblad_d1}) respectively.

\subsubsection*{Lamb-Dicke regime}

When the condition $\ev{(k_{\mu,z} z_j)^2}^{1/2} \ll 1$ is satisfied for every ion, the motion is in the Lamb-Dicke regime \cite{wineland1998JRNIST} and the master equation can be expanded in a series expansion in $k_{\mu,z} z_j$ \cite{morigi2003PRA}. In our setup, the angles $\pm \theta$ are such that the $z$-components $k_{\mu,z}$ are (a) not too large to cause beyond Lamb-Dicke regime dynamics, and (b) not too small that the cooling is weak. For any wavevector, it is useful to recast its coupling to the $z$-motion in terms of the Lamb-Dicke parameters associated with the drumhead modes as 
\begin{equation}
k_z z_j = \sum_{i=1}^N \eta_n^{k_z} \mathcal{M}_{jn} (b_n + b_n^\dag),
\end{equation}

\noindent where $\eta_n^{k_z} = k_z\sqrt{\frac{\hbar}{2M\omega_n}}$ is the Lamb-Dicke parameter \cite{wineland1998JRNIST} for mode $n$, associated with a wavevector whose $z$-component is $k_z$. Expanding the master equation Eq.~(\ref{eqn:master_eqn_full}) up to second-order in the Lamb-Dicke parameters results in
\begin{equation}
\dot{\rho} = \lv_0 \rho + \lv_1 \rho + \lv_2 \rho.
\label{eqn:master_eqn_second_order}
\end{equation}

Here,
\begin{eqnarray}
\lv_0 \rho &&= -i [H_0(t), \rho] + \sum_{\mu,j} \Gamma_\mu \mathcal{D}[\sigma_{g_\mu e}^j]\rho, \nonumber\\*
\lv_1 \rho &&= -i [H_1,\rho], \nonumber\\*
\lv_2 \rho &&= -i [H_2,\rho] + \mathcal{K}_2\rho, 
\end{eqnarray}

\noindent with 
%\begin{widetext}
\begin{eqnarray}
&&H_0(t) = \sum_n \omega_n b_n^\dag b_n + \sum_{j,\mu} \Delta_{\mu,j}(t) \sigma_{g_\mu g_\mu}^j  
 \nonumber\\* 
&&\hphantom{H_0(t) }+ \sum_{j,\mu} \frac{ \Omega_{\mu,j}}{2} \sigma_{eg_\mu}^j + \text{H.c.},\nonumber
\end{eqnarray}
\begin{eqnarray}
&&H_1 = \sum_{j,n,\mu} \frac{i \lambda_{jn}^{\mu}\Omega_{\mu,j}}{2} X_n \sigma_{eg_\mu}^j + \text{H.c.},\nonumber
\end{eqnarray}
\begin{eqnarray}
&&H_2 = -\sum_{j,n,k,\mu} \frac{\lambda_{jn}^{\mu} \lambda_{jk}^{\mu} \Omega_{\mu,j}}{4} X_n X_k\sigma_{eg_\mu}^j + \text{H.c.}, \nonumber
\end{eqnarray}
\begin{eqnarray}
&&\mathcal{K}_2\rho =\sum_{j,n,k,\mu} \frac{\Gamma_\mu}{2}\ev{u^2}_{eg_\mu} \lambda_{jn}^{\text{sc}} \lambda_{jk}^{\text{sc}} \times \nonumber\\*
&&\hphantom{\mathcal{K}_2\rho} \sigma_{g_\mu e}^j \left( 2 X_n \rho X_k -X_nX_k\rho -\rho X_n X_k \right) \sigma_{eg_\mu}^j,
\label{eqn:master_eqn_terms}
\end{eqnarray}

\noindent where $X_n = b_n + b_n^\dag$ and $\lambda_{jn}^{\mu} = \eta_n^{k_{\mu,z}}\mathcal{M}_{jn}$, $\lambda_{jn}^{\text{sc}} = \eta_n^{k_\text{sc}}\mathcal{M}_{jn}$ are dimensionless electronic-motional coupling strengths. The quantity $\ev{u^2}_{e g_\mu}$ is the variance of $u = \cos \theta_\text{sc}$ taken with respect to the dipole radiation pattern $\mathcal{N}_{e g_\mu}(u)$ associated with the $\ket{e}\rightarrow\ket{g_\mu}$ decay. The master equation, Eq.~(\ref{eqn:master_eqn_second_order}), is the starting point for our analysis of EIT cooling of ions in the Penning trap. 

\subsection{Parameters from the NIST EIT cooling experiment}

In the NIST EIT cooling experiment with ${}^9\text{Be}^+$ ions, the transverse magnetic field of $B = 4.46 \; \text{T}$ leads to a splitting of $124 \; \text{GHz}$ between the $2s^2 S_{1/2} (m_J = -1/2)$ and $2s^2 S_{1/2} (m_J = +1/2)$ levels \cite{biercuk2009QIC}, labeled as $\ket{g_1}$ and $\ket{g_2}$ respectively. The $\ket{g_1}\leftrightarrow\ket{e}$ transition frequency is $\omega_{g_1 e}/2\pi \approx 957 \; \text{THz}$. The two EIT lasers, with $\sigma^+$ and $\pi$ 
polarizations are oriented at $\pm 10^\circ$ with respect to the $x$-axis and respectively couple the $\ket{g_1}$ and $\ket{g_2}$ levels to the excited level $2p^2 P_{3/2} (m_J = +1/2)$, labeled as $\ket{e}$. They are blue detuned with equal detuning $\Delta^0$ from their respective transitions by hundreds of megahertz. These lasers generate sufficient power to give Rabi frequencies of tens of megahertz. The beam diameters ($\approx 1$ mm) of the EIT lasers are large compared to the diameters of the ion crystals ($\leq 300 \; \mu$m) so that we can assume constant laser intensities over the spatial extent of the crystal. The decay rates out of $\ket{e}$ are $\Gamma_1/2\pi \approx 6 \; \text{MHz}$ and $\Gamma_2/2\pi \approx 12 \; \text{MHz}$, with $\ev{u^2}_{e g_1} = 2/5$ and $\ev{u^2}_{e g_2} = 1/5$.

In all the calculations in this paper, we operate at the expected optimum EIT cooling condition for the COM mode of a stationary ion \cite{morigi2003PRA} given by $\Omega_1^2+\Omega_2^2 = 4\omega_{\COM}(\omega_{\COM}+\Delta^0)$, and assume equal Rabi frequencies, i.e. $\Omega_1 = \Omega_2 = \Omega_\text{opt}$, so that 

\begin{equation}
\Omega_\text{opt}(\Delta^0) = \sqrt{2\omega_\COM(\omega_\COM+\Delta^0)},
\label{eqn:opt_rabi}
\end{equation}  

\noindent where we use the value $\omega_\COM/2\pi = 1.59$ MHz. Further, we assume the rotation frequency of the crystal is $\omega_r/2\pi = 180$ kHz.

\section{\label{sec:single_ion} A single revolving ion}

\subsection{Time-varying Doppler shifts}

A toy model of a single ion revolving around the trap center in the $x$-$y$ plane can shed light on the impact of the in-plane motion on the cooling of the transverse motion. We recall that the circular in-plane motion of the ion causes a sinusoidally modulated Doppler shift, with the precise form of the modulation detailed in the paragraph immediately following Eq.~(\ref{eqn:int_Hamiltonian_1}). 

\begin{figure}[!htb]
\centering
\includegraphics[width=\linewidth]
{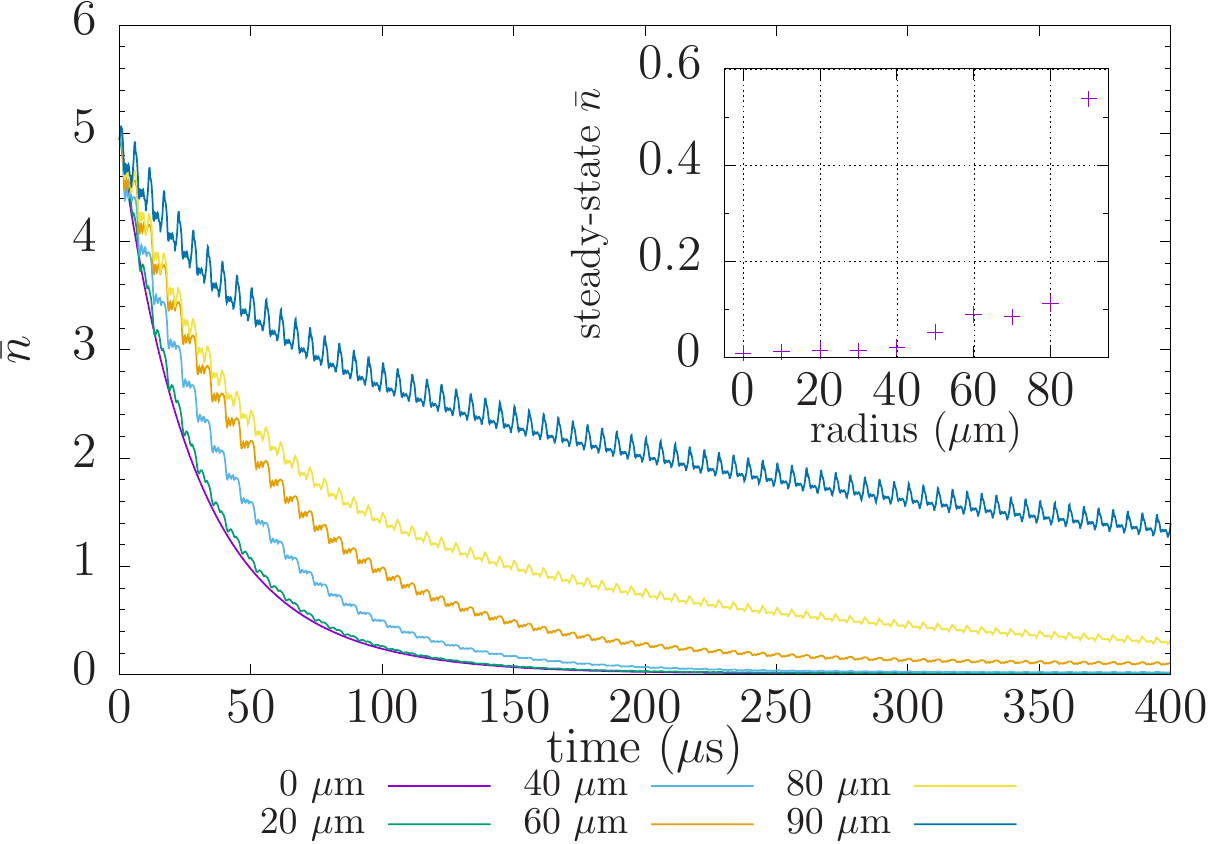}
\caption{(color online) Cooling of the transverse motion over time, for a single ion revolving around the trap center at different radii. The cooling is slower at larger radii because of time-varying Doppler shifts modulating the detunings of the EIT lasers as seen by the ion. Inset: The steady-state occupation also increases with distance from the trap center. Here, $\Delta^0/2\pi = 180$ MHz, $\Omega_1/2\pi=\Omega_2/2\pi=\Omega_\text{opt}(\Delta^0)/2\pi \approx 24$ MHz.}
\label{fig:single_ion_qutip_diff_radius}
\end{figure}

We set the detuning $\Delta^0/2\pi \approx 180 \; \text{MHz}$, and operate with equal Rabi frequencies $\Omega_\text{opt}(\Delta^0)$ given by Eq.~(\ref{eqn:opt_rabi}). We assume that a preceding Doppler cooling stage initializes the transverse motion of the ion to a thermal state with $\nbar = 5$. Typically, EIT cooling is applied after initializing the ion(s) in $\ket{g_1}$ by optical pumping. Figure~\ref{fig:single_ion_qutip_diff_radius} shows the decrease in the thermal occupation $\nbar$ with time as the EIT lasers address a single revolving ion, for different distances of the ion from the trap center. For a detuning $\sim 180 \; \text{MHz}$, the ion experiences effective red detunings for parts of its trajectory for a radius $r \gtrsim 50 \; \mu\text{m}$. Consequently, the ion undergoes heating in these regions, leading to slower cooling and higher final occupancies at larger radii (see inset of Fig. \ref{fig:single_ion_qutip_diff_radius}). Therefore, sufficiently large detunings have to be used, so that ions at the outer boundary of large crystals still experience an effective blue detuning of the EIT lasers.

\subsection{Timescale for internal dynamics} 

\begin{figure}[!htb]
\centering
\includegraphics[width=\linewidth]
{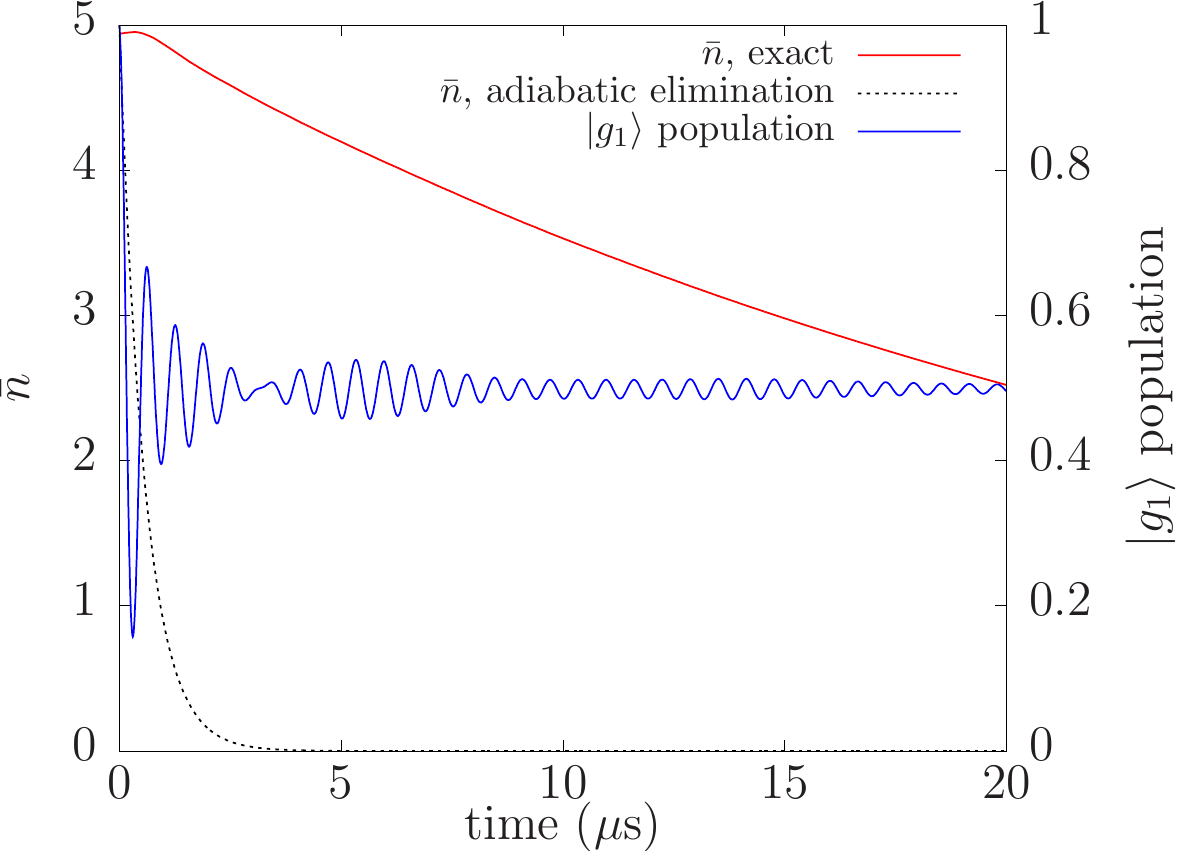}
\caption{(color online) Cooling curve for an ion at the trap center that experiences no Doppler shifts. The separation of the timescales for the electronic and motional degrees of freedom is not large enough to validate the adiabatic elimination of the electronic degrees of freedom. The cooling is therefore much slower than the result predicted from such an elimination procedure \cite{morigi2003PRA}. Here, $\Delta^0/2\pi = 180$ MHz, $\Omega_1/2\pi=\Omega_2/2\pi=\Omega_\text{opt}(\Delta^0)/2\pi \approx 24$ MHz.}
\label{fig:single_ion_qutip_internal_external_timescale}
\end{figure}

An ion located at the trap center experiences no Doppler shifts, and hence, we could argue that analytical expressions derived elsewhere \cite{morigi2003PRA} for EIT cooling of a single ion might be valid in such a situation. With the EIT wavevectors making an angle of $\pm 10^\circ$ with the $x$-axis, the Lamb-Dicke parameters are $\eta^{k_{1,z}} \approx -\eta^{k_{2,z}} \approx 0.066$. Combined with the typical Rabi frequencies used in the experiment, in the range of tens of megahertz, the wide separation of electronic and motion timescales demanded by an adiabatic elimination procedure is not satisfied in our system \cite{morigi2003PRA}. Figure~\ref{fig:single_ion_qutip_internal_external_timescale} shows the disagreement between the cooling curves obtained with (black dashed line) and without (red solid line) adiabatic elimination of the electronic degrees of freedom (DOF) for an ion at the trap center. The insufficient separation of timescales can also be seen qualitatively by simultaneously examining the transient dynamics of the population, in say, $\ket{g_1}$, on a common time axis, as shown in Fig.~\ref{fig:single_ion_qutip_internal_external_timescale}. 

\section{\label{sec:multi_ion} EIT cooling of multiple ions}

The NIST Penning trap routinely stores and manipulates tens to hundreds of ions. Since the density matrix, now consisting of the electronic degrees of freedom of all the ions and their drumhead modes, scales exponentially in the ion number, exact solutions are impossible, and we are forced to resort to approximate techniques. 

From the master equation, Eq. (\ref{eqn:master_eqn_second_order}), we write down the equations of motion for the means of all the system operators (first order moments) and the products of operator pairs (second order moments). In general, these equations will couple to higher order moments, for example, means of products of triplets of operators, and so on. We truncate the hierarchy at second order by neglecting all cumulants higher than means and covariances, and close the set of equations by approximating higher order moments using sums of products of first and second order moments \cite{meiser2010PRA,xu2015PRL}. As an example, for a product of three operators this would imply 
\begin{equation}
\ev{ABC} \approx \ev{AB}\ev{C} + \ev{AC}\ev{B} + \ev{BC}\ev{A} - 2 \ev{A}\ev{B}\ev{C}.
\end{equation}

Here $\ev{\ldots}$ denotes the mean value of an operator or product of operators. We will refer to the equations obtained for the first and second order moments using this truncation scheme as the Gaussian model (GM), since the scheme neglects third and higher order cumulants. We note, however, that we factorize second order moments involving the electronic degrees of freedom of different ions of the type $\ev{\sigma_{\alpha\beta}^j \sigma_{\gamma\delta}^k}$ as $\ev{\sigma_{\alpha\beta}^j} \ev{\sigma_{\gamma\delta}^k}$ for $j\neq k$. The equations arising from this cumulant expansion approach are detailed in Appendix \ref{app:eom}. 

\subsection{Results from the Gaussian model}

\begin{figure}[!htb]
\centering
\includegraphics[width=\linewidth]
{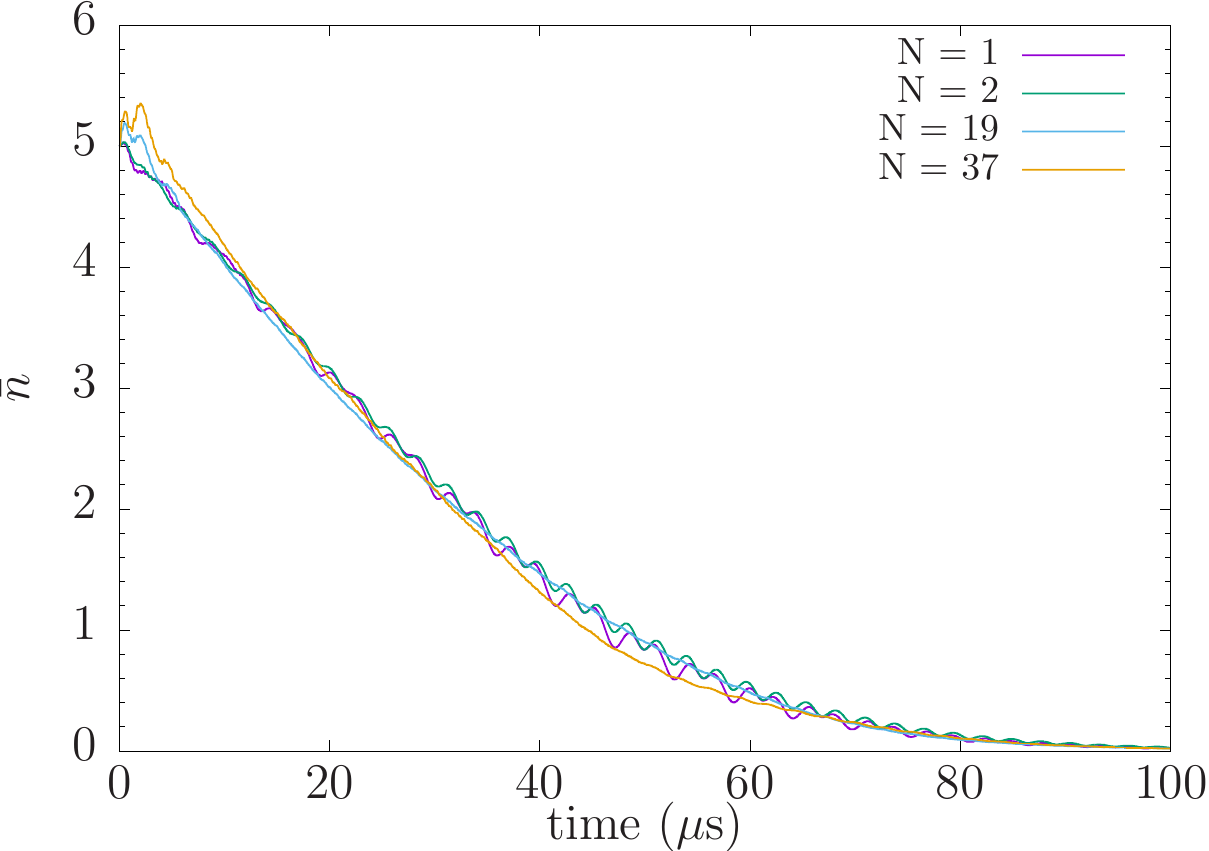}
\caption{(color online) Cooling curves for the center-of-mass (COM) mode for crystals with different ion numbers, calculated using the Gaussian model (GM), showing rapid near ground-state cooling within 100 $\mu$s. However, the cooling rates are almost identical for all of these crystals. Here, $\Delta^0/2\pi = 360$ MHz, $\Omega_1/2\pi=\Omega_2/2\pi=\Omega_\text{opt}(\Delta^0)/2\pi \approx 33.9$ MHz.}
\label{fig:cum_diff_N}
\end{figure}

We set the detuning $\Delta^0/2\pi = 360 \; \text{MHz}$, which ensures that the cooling rate for a single revolving ion does not change appreciably over the spatial extent of the small crystals we consider here. Figure~\ref{fig:cum_diff_N} shows the cooling of the COM mode for crystals with $N = 1,2,19$ and $37$ ions. For $N = 1, 2$ we simply take the ion(s) to be revolving at a distance of $20 \; \mu$m from the trap center, and diametrically opposite each other in the $N = 2$ case. In the case of multi-ion crystals $(N > 2)$, the equilibrium configuration of the crystal and the mode frequencies and eigenvectors are solved for following the procedure in Ref. \cite{wang2013PRA}. We assume that a preceding Doppler cooling stage initializes the COM mode to a thermal state with $\nbar = 5$. We choose the initial $\nbar$ of the remaining drumhead modes assuming that they are initially in thermal equilibrium with the COM mode. Qualitatively, the COM mode rapidly cools to near ground-state occupations within 100 microseconds. However, the cooling curves for the different crystals are nearly identical, showing that the net cooling rate of the COM mode, within the Gaussian framework, is approximately independent of the number of ions. 

\subsection{Benchmarking the Gaussian Model: Single-ion results}

\begin{figure}[!htb]
\centering
\includegraphics[width=\linewidth]
{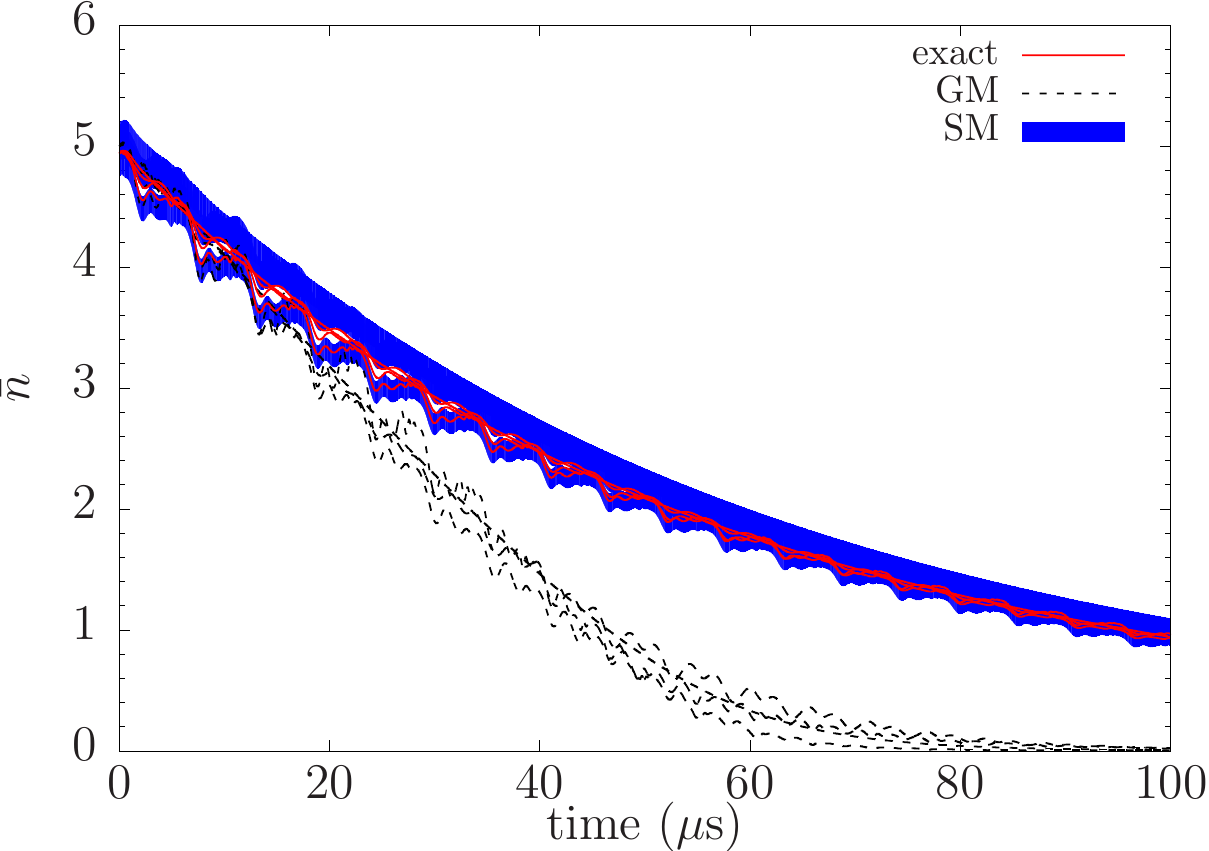}
\caption{(color online) Cooling curve for a single revolving ion, located at $r = 0,20,40$ and $60 \; \mu$m from the trap center, computed using three numerical approaches: (i) exact time evolution of the density matrix, (ii) the Gaussian model (GM) and (iii) the Sampling model (SM) using 2096 trajectories. The cooling curves from the GM do not agree with the exact curves. However, sampling the initial noise systematically (SM) accounts for beyond-Gaussian properties of the phase-space distribution of the system degrees of freedom, and reproduces the exact curves very well. Here, $\Delta^0/2\pi = 360$ MHz, $\Omega_1/2\pi=\Omega_2/2\pi=\Omega_\text{opt}(\Delta^0)/2\pi \approx 33.9$ MHz.}
\label{fig:N1_diff_methods}
\end{figure}

Since we are able to numerically solve the single-ion case exactly, we proceed to compare the results from the GM with the exact single-ion results. Figure~\ref{fig:N1_diff_methods} shows the cooling curves from the GM (black, dashed) and the exact computation (red, solid) for a single revolving ion at $r = 0,20, 40$ and $60 \; \mu\text{m}$ from the trap center, with $\Delta^0/2\pi = 360 \; \text{MHz}$. While both models qualitatively indicate that the cooling rate is roughly the same at these different radii, the cooling rate obtained from the GM is quantitatively very different from the exact result. This indicates that keeping track of only means and covariances is not sufficient to accurately capture the cooling curve. 

The GM assumes that in the combined phase space of all the system degrees of freedom (DOF), the joint (quasi)-probability distribution of these DOF remains Gaussian at all times. In reality, while the initial distribution is Gaussian, evolution under the subsequent dynamics generally distorts the distribution so that it is no longer Gaussian at later times. A systematic way to capture this effect is to construct moments by averaging the evolution of the corresponding phase space variables (or products of variables) over a large number of trajectories, where the initial conditions of these variables in each trajectory are chosen randomly from their initial distribution (see Table~\ref{tab:gaussian_vs_sampling}). Such a sampling and averaging procedure captures the build-up of non-trivial third and higher order cumulants that are neglected in the GM. We note that this approach is in the same spirit as the Truncated Wigner Approximation (TWA) used in calculating the dynamics of spin-spin and spin-boson models \cite{polkovnikov2010AnnPhys,blakie2008AdvPhys,schachenmayer2015PRX,orioli2017PRA}. Moreover, we track separate phase space variables corresponding to system operators as well as operator pairs, and evolve these variables using the same equations of motion as in the GM (Appendix~\ref{app:eom}), but for many trajectories. We thereby perform beyond mean-field calculations \cite{orioli2017PRA} that capture those contributions to the covariances between system operators which develop as a result of the subsequent diffusive-dissipative dynamics. We refer to this method as the Sampling model (SM). Table~\ref{tab:gaussian_vs_sampling} summarizes the implementational differences between the Gaussian and the beyond-Gaussian Sampling model.  
\begin{table}[!htb]
\begin{tabular}{ccc}
Model	&	Trajectories	&	Initial condition \\\hline\\
Gaussian (GM)	&	1	&	\makecell{$\ev{b_n}(0)=0$,\\ $\ev{b_n^\dag b_n}(0)=\nbar_n (0)$}\\\\\hline\\
Sampling (SM)	&	Many	&	\makecell{$\text{Re}\{\ev{b_n}(0)\} = \text{Gaus}(0,\sqrt{\nbar_n/2})$, \\$\text{Im}\{\ev{b_n}(0)\} = \text{Gaus}(0,\sqrt{\nbar_n/2})$, \\$\ev{b_n^\dag b_n(0)} =\abs{\ev{b_n}(0)}^2$}\\ 
\end{tabular}  
\caption{\label{tab:gaussian_vs_sampling} Implementational differences between the Gaussian and Sampling models. Here $\text{Gaus}(0,\sigma)$ is a Gaussian distributed random variable with zero mean and standard deviation $\sigma$, and $\nbar_n$ are the initial thermal mode occupations. In the sampling model, the quantity $\ev{A}$ simply denotes the value of the respective phase space variable in that trajectory, and is \emph{not} the mean value of the operator $A$. Instead, the mean value of $A$ is given by the average of $\ev{A}$ over many trajectories with random initial conditions drawn from the initial phase space distribution. We note that we only sample the initial thermal distribution of the normal modes, and initialize the electronic DOF in the same way as in the GM, i.e. $\ev{\sigma_{\alpha \beta}^j}=1$ when $\alpha=\beta=g_1$. } 
\end{table}

In Fig.~\ref{fig:N1_diff_methods} we show the cooling curves from the SM along with the exact as well as GM results for the single revolving ion. Since the SM involves averaging over multiple trajectories with randomly drawn initial conditions, the cooling curves are shown as $1 \text{-} \sigma$ confidence intervals instead of a line plot. The SM cooling curves agree very well with the exact result, indicating that beyond-Gaussian correlations develop during the cooling process that lower the cooling rate. 

\subsection{Results from the Sampling model} 

\begin{figure}[!htb]
\centering
\includegraphics[width=\linewidth]
{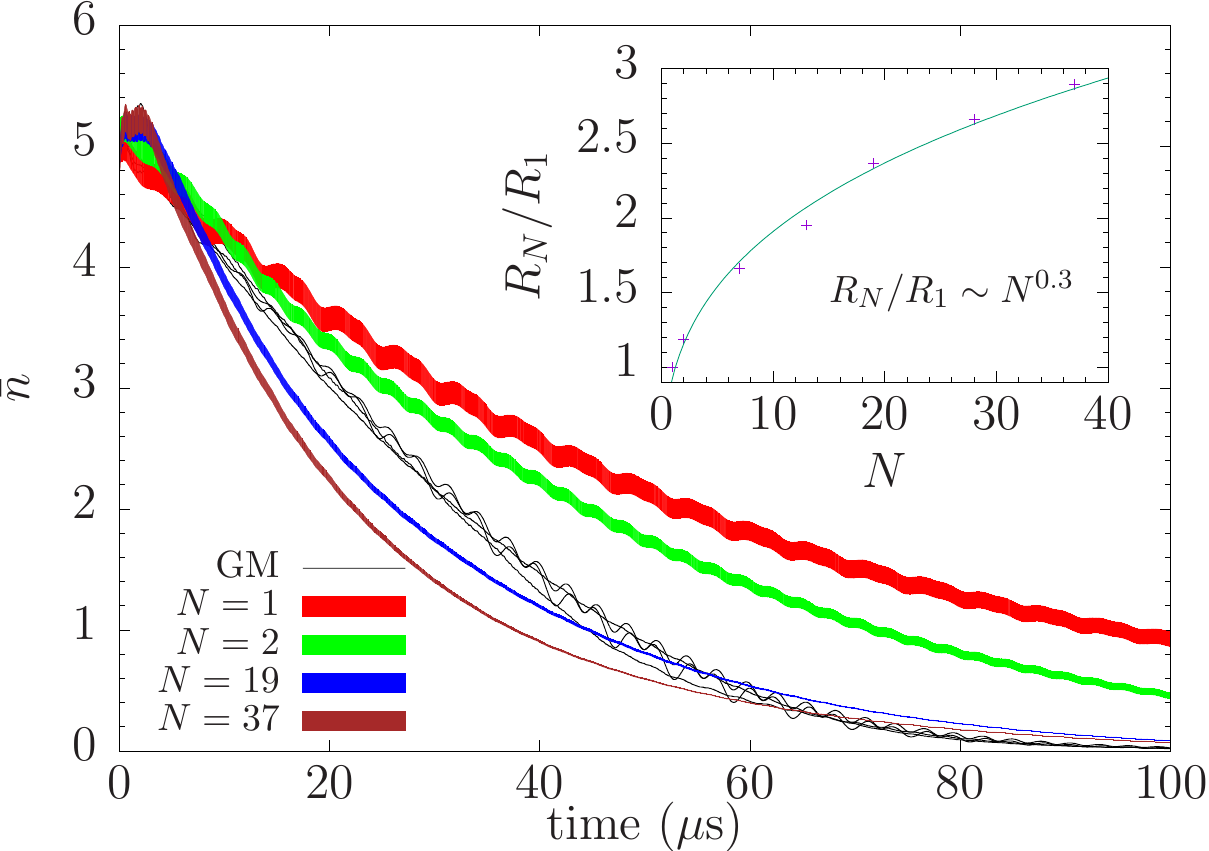}
\caption{(color online) Cooling curves for the center-of-mass (COM) mode for crystals with different ion numbers, calculated using the Sampling model (SM). The SM predicts that the cooling rate increases with ion number $N$. The cooling curves from the GM (Fig.~\ref{fig:cum_diff_N}) are also shown for comparison, where the $N$-dependency of the cooling rate does not manifest. Inset: Cooling rate of an $N$-ion crystal $R_N$ relative to the single-ion rate $R_1$ extracted from the SM (markers). A power-law fit (solid line) shows that the cooling rate scales as $\sim N^{0.3}$ for the parameters used. Here, $\Delta^0/2\pi = 360$ MHz, $\Omega_1/2\pi=\Omega_2/2\pi=\Omega_\text{opt}(\Delta^0)/2\pi \approx 33.9$ MHz.}
\label{fig:traj_diff_N}
\end{figure}

The SM predicts that the cooling rate of the COM mode increases with the number of ions $N$ in the crystal. In Fig.~\ref{fig:traj_diff_N}, we plot the cooling curve for the COM mode for crystals with $N=1,2,19$ and $37$ ions. The cooling is faster in the $N=2$ case than in the single ion case, and even faster in the $19$ and $37$ ion crystals. The inset shows the ratio of the cooling rate $R_N$ of an $N$-ion crystal to the rate $R_1$ for a single ion. With the parameters used, the cooling rate scales as $\sim N^{0.3}$, highlighting that EIT cooling of multiple ions cannot be explained trivially as the net cooling resulting from the individual ions. Rather, the $N$-dependency of the cooling rate indicates the intrinsic many-body nature of this problem. Further, the rapid nature of EIT cooling is evident from the time constant $\tau \approx 21 \; \mu$s for the cooling curve of the $37$ ion crystal. 

\section{\label{sec:full_bandwidth}Cooling over the full bandwidth}

\begin{figure}[!htb]
\centering
\includegraphics[width=\linewidth]
{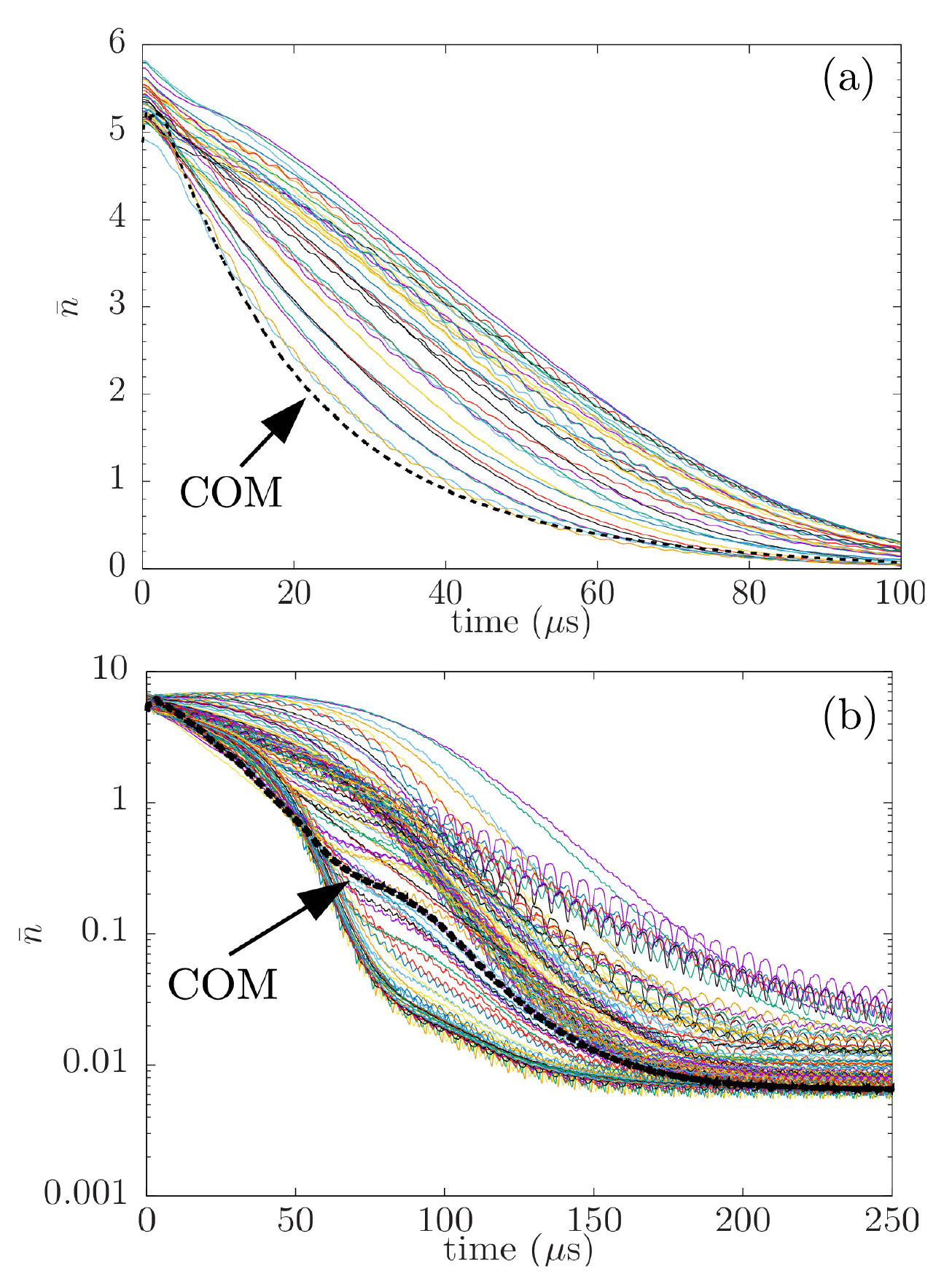}
\caption{(color online) (a) Cooling curves for all the drumhead modes of a 37-ion crystal with bandwidth $B/2\pi \approx 185$ kHz, computed using the SM, showing efficient cooling within 100 $\mu$s. Here, $\Delta^0/2\pi = 360$ MHz, $\Omega_1=\Omega_2=\Omega_\text{opt}(\Delta^0) \approx 33.9$ MHz. (b) GM cooling curves ($y$-axis in logscale) for all the drumhead modes of a 120-ion crystal with bandwidth $B/2\pi \approx 376$ kHz, showing near ground-state, steady-state occupations of all the modes after few hundred microseconds of EIT cooling. Here, $\Delta^0/2\pi = 400$ MHz, $\Omega_1/2\pi=\Omega_2/2\pi=\Omega_\text{opt}(\Delta^0)/2\pi \approx 35.7$ MHz. The same rotating wall frequency, $\omega_r/2\pi = 180$ kHz was used in both cases.}
\label{fig:full_bandwidth}
\end{figure}

The full bandwidth of drumhead modes are typically cooled to near ground-state occupancies in a 
single experimental application of EIT cooling with a fixed set of parameters. In Fig.~\ref{fig:full_bandwidth}(a), we show the cooling transients for all the tranverse modes of a 37-ion crystal, calculated using the SM. The bandwidth $B$ of the drumhead modes is $B/2\pi \approx 185$ kHz. All the modes are observed to reach near ground-state occupations within 100 microseconds. 

The computational complexity of a single trajectory in the SM scales as $N^3$ with the number of ions $N$, thereby making trajectory computations for large crystals ($\gtrsim 60$) untractable. However, the GM and SM will result in the same steady-state results since they only differ in the initial conditions, and eventually dissipation leads to the system of equations losing memory of its initial conditions. Therefore, the GM can be used to study the final temperatures that result from EIT cooling of large ion crystals, as shown in Fig.~\ref{fig:full_bandwidth}(b) for a 120-ion crystal. From initial occupations in the range $\nbar \approx 5 - 7$, all the modes are cooled down to $\nbar < 0.1$, showing the efficient cooling over the full bandwidth of drumhead modes, which in this case is $B/2\pi \approx 376$ kHz. We note that the experimentally observed occupations are expected to be somewhat higher than the steady-state values attained here because of the approximate model used in the simulations. Although the GM cooling transients are not completely reliable, they nevertheless indicate that a few hundred microseconds of EIT cooling is sufficient to achieve these very low occupations.     

\section{\label{sec:misalign} Sensitivity to laser alignment}

In modeling the EIT cooling in Sections~\ref{sec:single_ion}, \ref{sec:multi_ion} and \ref{sec:full_bandwidth}, we have assumed that the lasers are perfectly aligned, i.e. $k_{2,x} = k_{1,x}$, so that their difference wavevector is along the $z$-axis. In practice, a small misalignment of the EIT lasers could result in a component of the difference wavevector along the in-plane $x$-axis, because $k_{2,x} \neq k_{1,x}$. As a result, the detunings of these dressing lasers are now modulated unequally by Doppler shifts arising from the large amplitude in-plane rotation of the ion crystal, so that the instantaneous detunings of the two lasers as seen by the ion, $\Delta_{\mu}(t) = \Delta_{\mu}^0 - k_{\mu,x}v_x(t)$ with $\mu = 1,2$, are no longer identical. 

We study the effect of such a misalignment by considering a single ion revolving at different distances $r$ from the trap center. We introduce a small misalignment $\delta \theta$ that modifies the perfectly aligned $\mbf{k}_2 \rightarrow \mbf{k}_2^\text{(m)}$ such that,
\begin{eqnarray}
k_{2,x}^\text{(m)} &=& k_{2,x}\cos\delta\theta + k_{2,z}\sin\delta\theta \\\nonumber
k_{2,z}^\text{(m)} &=& k_{2,z}\cos\delta\theta - k_{2,x}\sin\delta\theta,
\end{eqnarray}  

\noindent where the subscript (m) denotes the misaligned $\mbf{k}_2$ vector.

\begin{figure}[!htb]
\centering
\includegraphics[width=\linewidth]
{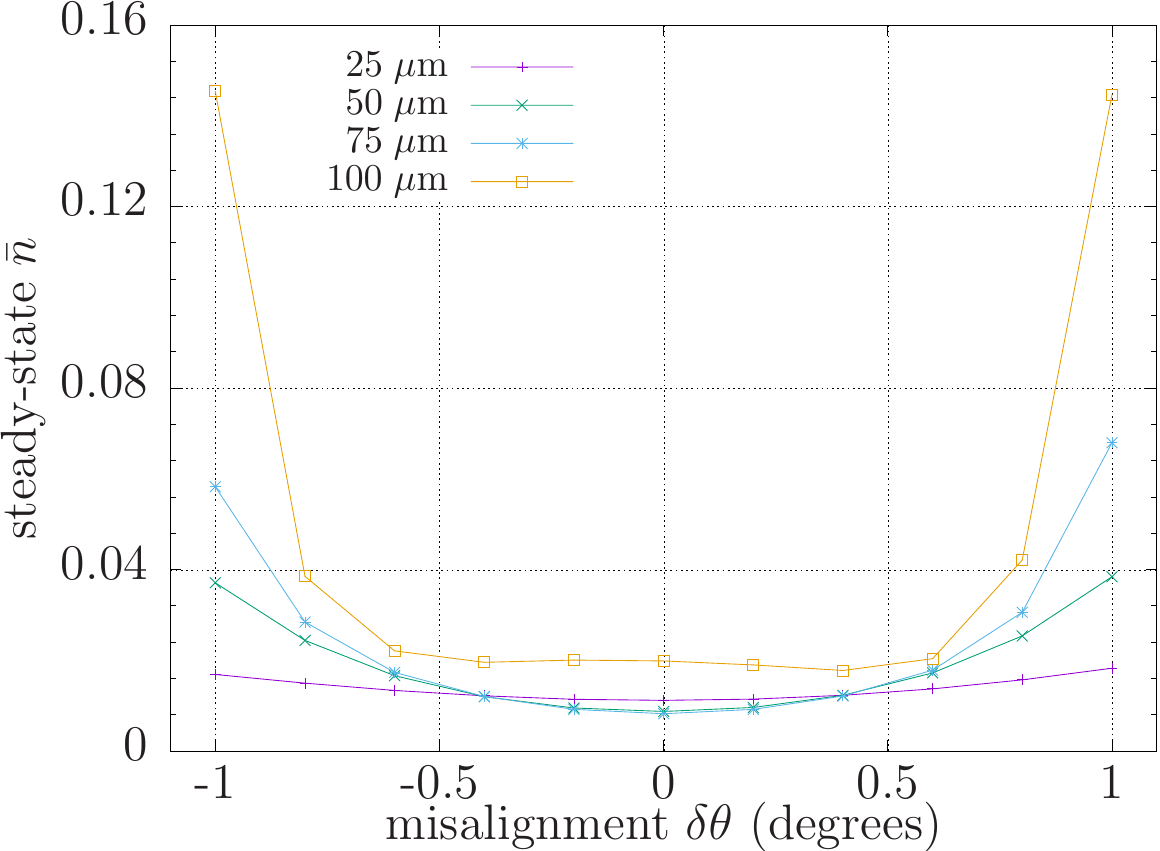}
\caption{(color online) Steady-state occupation as a function of the misalignment angle $\delta \theta$ for a single ion revolving around the trap center at different radii. The final occupation is not very sensitive to small misalignments ($\delta \theta \leq 1^\circ$) of the EIT wavevectors.  Here, $\Delta^0/2\pi = 400$ MHz, $\Omega_1/2\pi=\Omega_2/2\pi=\Omega_\text{opt}(\Delta^0)/2\pi \approx 35.7$ MHz.}
\label{fig:misalign}
\end{figure}

Figure~\ref{fig:misalign} shows the final steady-state occupation $\bar{n}$ of a single ion revolving around the trap center at different radii $r$, as the misalignment angle $\delta \theta$ is varied. We set the detuning $\Delta^0/2\pi = 400$ MHz, which ensures that for $r \leq 110 \; \mu$m, the revolving ion experiences effective blue detuning throughout its trajectory. The steady-state $\bar{n}$ begins to significantly increase only when $\lvert \delta \theta \rvert \gtrsim 1^\circ$, at which point the plot still indicates $\bar{n} \ll 1$. As $r$ varies from $25 \; \mu$m to $75 \; \mu$m, the sensitivity to $\delta \theta$ also increases with $r$. For $r = 100 \; \mu$m, the ion experiences only small blue detunings for parts of its trajectory, resulting in inefficient cooling in these regions even with perfect alignment. As a result, the final temperature is not very sensitive to small misalignments such that $\lvert \delta \theta \rvert \lesssim 0.5^\circ$. However, the final temperature grows sharply as $\lvert \delta \theta \rvert$ increases beyond this value.

In the NIST EIT cooling experiment, the misalignment between the two EIT wavevectors was estimated to be $< 0.2^\circ$ \cite{jordan2018}, ensuring that the cooling is negligibly affected by the in-plane crystal rotation. We note, however, that our analysis of the laser misalignment does not consider the potential adverse effect of the EIT lasers on the in-plane modes, which in turn could degrade the cooling of the drumhead modes. Such an analysis might result in a more stringent restriction on the tolerable range of $\delta \theta$.

\section{\label{sec:conclusion} Conclusion}

Our numerical study shows that EIT cooling is a robust technique for cooling all the drumhead modes of two-dimensional ion crystals in Penning traps to near ground-state occupancies. EIT cooling relies on quantum interference effects for its operation, and prior to our work, the chances for its success, dependent on delicate cancellations of absorption amplitudes, in a challenging setting such as a Penning trap were very uncertain. Multiple factors could have potentially led to the failure of EIT cooling, namely, Doppler shifts, insufficient separation of electronic and motional timescales, as well as simultaneous cooling of multiple ions. Our predictions for the success of EIT cooling have been validated by the successul experimental demonstration of EIT cooling of crystals with more than 100 ions in the NIST Penning trap \cite{jordan2018}. Quantitative measurements of the cooling rate and final occupation of the COM mode, as well as qualitative features observed over the full bandwidth of modes are consistent with the expectations from our numerical study. These theoretical and experimental results highlight the robustness of EIT cooling and make it an attractive scheme to cool large chains or arrays of trapped ions in other settings \cite{pagano2018arxiv,richerme2016PRA}.  

In the future, an important aspect to investigate is the effect of higher-order anharmonic terms in the trap potential, and more importantly, in the Coulomb interaction. These anharmonic terms not only result in additional coupling of the different drumhead modes to each other, but also couple these modes to the thermal motion associated with the in-plane modes. An understanding of these anharmonic couplings, at least for small ion crystals, will provide great insight into whether EIT cooling of the drumhead modes can also indirectly cool the in-plane modes, and also conversely, how the temperature of the in-plane modes could limit the achievable drumhead mode temperatures as well as cooling rates. Finally, while we have numerically observed and also found experimental support for the enhancement in the cooling rate with the number of ions, an intuitive explanation for this surprising feature will greatly illuminate the role played by many-body physics in the cooling dynamics.

\section*{Acknowledgments}

We thank Y. Lin, M. Affolter, J. Cooper and R.J. Lewis-Swan for stimulating discussions. We acknowledge the use of the Quantum Toolbox in Python (QuTiP) \cite{johansson2012CompPhys,johansson2013CompPhys} for the exact single-ion numerical calculations presented in this paper. This work was supported by NSF grants PHY 1734006 and PHY 1820885, DARPA Extreme Sensing, the Air Force Office of Scientific Research grants FA9550-18-1-0319 and its Multidisciplinary University Research Initiative grant (MURI), Army Research Office grant W911NF-16-1-0576, JILA-NSF grant PFC-173400, and NIST. JEJ gratefully acknowledges the Leopoldina Research Fellowship, German National Academy of Sciences grant LPDS 2016-15, and the NIST-PREP program. This manuscript is a contribution of NIST and not subject to U.S. copyright.

%\clearpage
\appendix

%use regular section syntax here
\section{\label{app:eom}Equations of motion for first and second order moments}

We classify the moments into three categories: internal, external and hybrid moments. These are listed in Table \ref{tab:list_moments}. For brevity, here we write down only the equations of motion (EOM) for the moments not marked with a \# in Table \ref{tab:list_moments}, from which the remaining EOM can be obtained by exchanging $g_1\leftrightarrow g_2$. 

\begin{table}[!htb]
\begin{tabular}{l@{\hspace{2pc}}c}
Type	&	Moments\\\hline\\
Internal	&	\makecell[l]{$\ev{\sigma_{g_1 g_1}^j}$, $\ev{\sigma_{g_1 g_2}^j}$, $\ev{\sigma_{g_1 e}^j}$, \\ $\ev{\sigma_{g_2 g_2}^j}^\#$, $\ev{\sigma_{g_2 e}^j}^\#$} \hspace*{\fill}\\\\
External	&	$\ev{b_n}$, $\ev{b_n b_k}$, $\ev{b_n^\dag b_k}$ \hspace*{\fill}\\\\
Hybrid	&	\makecell[l]{$\ev{b_n \sigma_{g_1 g_1}^j}$, $\ev{b_n \sigma_{g_1 g_2}^j}$, $\ev{b_n \sigma_{g_1 e}^j}$ \\ 
$\ev{b_n \sigma_{g_2 g_1}^j}^\#$, $\ev{b_n \sigma_{g_2 g_2}^j}^\#$, $\ev{b_n \sigma_{g_2 e}^j}^\#$ \\
$\ev{b_n \sigma_{e g_1}^j}$, $\ev{b_n \sigma_{e g_2}^j}^\#$}\hspace*{\fill}
\end{tabular}
\caption{\label{tab:list_moments} List of moments classified according to the nature of the operators involved. The equations for the moments marked with a \# can be derived trivially by exchanging $g_1\leftrightarrow g_2$ in the appropriate equations of motion for the other moments.} 
\end{table}

In Table \ref{tab:partial_sums}, we introduce a set of ``partial sums" that not only simplify the notation, but also speed up the computation time by identifying recurring summations and evaluating them only once per time step.

\begin{table}[!htb]
\begin{tabular}{l@{\hspace{2pc}}c}
Symbol	&	Definition\\\hline\\
$\mP_{\mu,j}^X$	&	$\sum_m \lambda_{jm}^\mu \ev{X_m}$ \hspace*{\fill}\\[5pt]
$\mP_{\mu,j}^{X X}$	&	$\sum_{l,m} \lambda_{jl}^\mu \lambda_{jm}^\mu \ev{X_l X_m}$ \hspace*{\fill}\\[5pt]
$\mP_{\mu,jn}^{b X}$	&	$\sum_m \lambda_{jm}^\mu \ev{b_n X_m}$ \hspace*{\fill}\\[5pt]
$\mP_{\mu,jn}^{d X}$	&	$\sum_m \lambda_{jm}^\mu \ev{b_n^\dag X_m}$ \hspace*{\fill}\\[5pt]
$\mP_{\mu,j}^{X \sigma_\alpha}$	&	$\sum_m \lambda_{jm}^\mu \ev{X_m \sigma_\alpha^j}$ \hspace*{\fill}\\[5pt]
$\mP_{\mu,[qj]}^{X \sigma_\alpha}$	&	$\sum_m \lambda_{qm}^\mu \ev{X_m \sigma_\alpha^j}$\hspace*{\fill} 
\end{tabular}
\caption{\label{tab:partial_sums} Definition of partial sums to simplify notation and speed up computation.} 
\end{table}

\begin{widetext}

\subsection{Internal moments}

\begin{eqnarray}
\frac{d}{dt}\ev{\sigma_{g_1g_1}^j} &=& -\frac{i}{2}\left( \Omega_{1,j}^*\ev{\sigma_{g_1e}^j} - \; \text{c.c.} \right) + \Gamma_1 \left(1 - \ev{\sigma_{g_1g_1}^j} - \ev{\sigma_{g_2g_2}^j}\right)
-\frac{1}{2}\left( \Omega_{1,j}^*\mP_{1,j}^{X \sigma_{g_1e}}  + \; \text{c.c.} \right) \nonumber\\
%&&\hphantom{\frac{d}{dt}\ev{\sigma_{g_1g_1}^j} = }
&&{}
+ \frac{i}{4} \left\{ \Omega_{1,j}^*\left( \mP_{1,j}^{X X}\ev{\sigma_{g_1e}^j} + 2 \mP_{1,j}^X \mP_{1,j}^{X \sigma_{g_1e}} - 2 (\mP_{1,j}^X)^2 \ev{\sigma_{g_1e}^j} \right) - \; \text{c.c.} \right\} 
\end{eqnarray}

\begin{eqnarray}
\frac{d}{dt}\ev{\sigma_{g_1g_2}^j} &=& 
i\left(\Delta_{1,j}(t)-\Delta_{2,j}(t)\right)\ev{\sigma_{g_1g_2}^j} 
+ \frac{i\Omega_{1,j}}{2}\ev{\sigma_{g_2e}^j}^* 
- \frac{i\Omega_{2,j}^*}{2}\ev{\sigma_{g_1e}^j}
-\frac{\Omega_{1,j}}{2}\mP_{1,j}^{X \sigma_{eg_2}} 
-\frac{\Omega_{2,j}^*}{2}\mP_{2,j}^{X \sigma_{g_1e}} \nonumber\\
%&&\hphantom{\frac{d}{dt}\ev{\sigma_{g_1g_2}^j}  }
&&{}-\frac{i\Omega_{1,j}}{4}\left( \mP_{1,j}^{X X}\ev{\sigma_{g_2e}^j}^* 
+ 2 \mP_{1,j}^X \mP_{1,j}^{X \sigma_{eg_2}}
- 2 (\mP_{1,j}^X)^2 \ev{\sigma_{g_2e}^j}^* \right) \nonumber\\
%&&\hphantom{\frac{d}{dt}\ev{\sigma_{g_1g_2}^j}  }
&&{}+\frac{i\Omega_{2,j}^*}{4}\left( \mP_{2,j}^{X X}\ev{\sigma_{g_1e}^j} 
+ 2 \mP_{2,j}^X \mP_{2,j}^{X \sigma_{g_1e}}
- 2 (\mP_{2,j}^X)^2 \ev{\sigma_{g_1e}^j} \right)
\end{eqnarray}

\begin{eqnarray}
&&\frac{d}{dt}\ev{\sigma_{g_1e}^j} = 
-\left( \frac{\Gamma}{2}-i\Delta_{1,j}(t)\right)\ev{\sigma_{g_1e}^j}
-\frac{i\Omega_{1,j}}{2}\left(2\ev{\sigma_{g_1g_1}^j}+\ev{\sigma_{g_2g_2}^j}-1\right)
-\frac{i\Omega_{2,j}}{2}\ev{\sigma_{g_1g_2}^j} \nonumber\\
&&\hphantom{\frac{d}{dt}\ev{\sigma_{g_1e}^j} = }
+\frac{\Omega_{1,j}}{2} \left( 2 \mP_{1,j}^{X \sigma_{g_1g_1}} + \mP_{1,j}^{X \sigma_{g_2g_2}} - \mP_{1,j}^X \right)
+\frac{\Omega_{2,j}}{2} \mP_{2,j}^{X \sigma_{g_1g_2}} \nonumber\\
&&\hphantom{\frac{d}{dt}\ev{\sigma_{g_1e}^j} = }
+\frac{i\Omega_{1,j}}{4}\left( \mP_{1,j}^{X X}( 2\ev{\sigma_{g_1g_1}^j}+\ev{\sigma_{g_2g_2}^j}-1 ) 
+ 2 \mP_{1,j}^X ( 2\mP_{1,j}^{X \sigma_{g_1g_1}} + \mP_{1,j}^{X \sigma_{g_2g_2}} - \mP_{1,j}^X )  
- 2 (\mP_{1,j}^X)^2 ( 2\ev{\sigma_{g_1g_1}^j}+\ev{\sigma_{g_2g_2}^j}-1 ) \right) \nonumber\\
&&\hphantom{\frac{d}{dt}\ev{\sigma_{g_1e}^j} = }
+\frac{i\Omega_{2,j}}{4}\left( \mP_{2,j}^{X X} \ev{\sigma_{g_1g_2}^j}
+ 2 \mP_{2,j}^X \mP_{2,j}^{X \sigma_{g_1g_2}}
- 2 (\mP_{1,j}^X)^2 \ev{\sigma_{g_1g_2}^j} \right)
\end{eqnarray}

\subsection{External moments}

%\end{widetext}

In the following equations, the index $\mu$ takes on values $1,2$ to account for the two EIT lasers. 

%\begin{widetext}

\begin{eqnarray}
&&\frac{d}{dt}\ev{b_n} = -i\omega_n \ev{b_n} -\sum_{\mu,j} \frac{\lambda_{jn}^\mu}{2} \left( \Omega_{\mu,j}^* \ev{\sigma_{g_\mu e}^j} - \; \text{c.c.} \right)  
+\sum_{\mu,j} \frac{i\lambda_{j,n}^\mu}{2} \left( \Omega_{\mu,j}^* \mP_{\mu,j}^{X \sigma_{g_\mu e}} + \; \text{c.c.} \right)
\end{eqnarray}

\begin{eqnarray}
&&\frac{d}{dt}\ev{b_n b_k} = -i(\omega_n+\omega_k) \ev{b_n b_k} 
-\sum_{\mu,j} \Gamma_\mu \ev{u^2}_{eg_\mu} \lambda_{\mu,jn}^{\text{sc}} \lambda_{\mu,jk}^{\text{sc}} (1 - \ev{\sigma_{g_1g_1}^j} - \ev{\sigma_{g_2g_2}^j}) \nonumber\\
&& 
- \sum_{\mu,j} \frac{\lambda_{j,n}^\mu}{2} \left( \Omega_{\mu,j}^* \ev{b_k \sigma_{g_\mu e}^j} - \Omega_{\mu,j} \ev{b_k \sigma_{eg_\mu}^j}  \right) + \; n \leftrightarrow k  \nonumber\\
&&
+ \sum_{\mu,j} \frac{\lambda_{j,n}^\mu}{2} \left( \Omega_{\mu,j}^* 
\left\{ (\mP_{\mu,jk}^{d X})^* \ev{\sigma_{g_\mu e}^j} 
+ \mP_{\mu,j}^X \ev{b_k \sigma_{g_\mu e}^j} 
+ \mP_{\mu,j}^{X \sigma_{g_\mu e}} \ev{b_k}
- 2 \mP_{\mu,j}^X \ev{\sigma_{g_\mu e}^j} \ev{b_k} \right\} 
+ \Omega_{\mu,j}\left\{ \vphantom{\mP_{\mu,j}^{X \sigma_{g_\mu e}}} g_\mu e \rightarrow eg_\mu\right\}   
\right) + \; n \leftrightarrow k \nonumber\\
\end{eqnarray}

\begin{eqnarray}
&&\frac{d}{dt}\ev{b_n^\dag b_k} = -i(\omega_k-\omega_n) \ev{b_n^\dag b_k} +  \sum_{\mu,j} \Gamma_\mu \ev{u^2}_{eg_\mu} \lambda_{\mu,jn}^{\text{sc}} \lambda_{\mu,jk}^{\text{sc}} (1 - \ev{\sigma_{g_1g_1}^j} - \ev{\sigma_{g_2g_2}^j}) \nonumber\\
&&
+ \sum_{\mu,j} \frac{\lambda_{j,n}^\mu}{2} \left( \Omega_{\mu,j}^* \ev{b_k \sigma_{g_\mu e}^j} - \Omega_{\mu,j} \ev{b_k \sigma_{eg_\mu}^j}  \right) 
- \sum_{\mu,j} \frac{\lambda_{j,k}^\mu}{2} \left( \Omega_{\mu,j}^* \ev{b_n \sigma_{e g_\mu}^j}^* - \Omega_{\mu,j} \ev{b_n \sigma_{g_\mu e}^j}^*  \right) \nonumber\\
&&
- \sum_{\mu,j} \frac{\lambda_{j,n}^\mu}{2} \left( \Omega_{\mu,j}^* 
\left\{ (\mP_{\mu,jk}^{d X})^* \ev{\sigma_{g_\mu e}^j} 
+ \mP_{\mu,j}^X \ev{b_k \sigma_{g_\mu e}^j} 
+ \mP_{\mu,j}^{X \sigma_{g_\mu e}} \ev{b_k}
- 2 \mP_{\mu,j}^X \ev{\sigma_{g_\mu e}^j} \ev{b_k} \right\} 
+ \Omega_{\mu,j}\left\{ \vphantom{\mP_{\mu,j}^{X \sigma_{g_\mu e}}} g_\mu e \rightarrow eg_\mu \right\}   
\right) \nonumber\\
&&
+ \sum_{\mu,j} \frac{\lambda_{j,k}^\mu}{2} \left( \Omega_{\mu,j}^* 
\left\{ \mP_{\mu,jn}^{d X} \ev{\sigma_{g_\mu e}^j} 
+ \mP_{\mu,j}^X \ev{b_n \sigma_{e g_\mu}^j}^* 
+ \mP_{\mu,j}^{X \sigma_{g_\mu e}} \ev{b_n}^*
- 2 \mP_{\mu,j}^X \ev{\sigma_{g_\mu e}^j} \ev{b_n}^* \right\} 
+ \Omega_{\mu,j}\left\{ \vphantom{\mP_{\mu,j}^{X \sigma_{g_\mu e}}} g_\mu e \rightarrow eg_\mu \right\}   
\right)
\end{eqnarray}

\clearpage

\end{widetext}

\subsection{Hybrid moments}

Table \ref{tab:ions_partial_sums} introduces some additional partial sums, now over the ions instead of the modes, that will further aid in compact presentation and faster computation by identification of recurring summations. 

\vspace*{5pt}

\begin{table}[!htb]
\begin{tabular}{l@{\hspace{1pc}}c}
Symbol	&	Definition\\\hline\\
$\mQ_{jn}^{\lv_1}$	&	$-\sum_{\mu,q \neq j} \frac{\lambda_{q,n}^\mu}{2} 
\left( \Omega_{\mu,q}^* \ev{\sigma_{g_\mu e}^q} -\; \text{c.c.}\right)$\hspace*{\fill} \\[10pt]
$\mQ_{jn}^{\lv_2 (1),\sigma_\alpha}$	&	$\sum_{\mu,q \neq j} \frac{i \lambda_{qn}^\mu}{2} \mP_{\mu,[qj]}^{X \sigma_\alpha} \left( \Omega_{\mu,q}^* \ev{\sigma_{g_\mu e}^q} + \; \text{c.c.} \right)$ \hspace*{\fill}\\[10pt]
$\mQ_{jn}^{\lv_2 (2)}$	&	$\sum_{\mu,q \neq j} \frac{i \lambda_{qn}^\mu}{2} \left( \Omega_{\mu,q}^* \mP_{\mu,q}^{X \sigma_{g_\mu e}} + \; \text{c.c.}\right)$ \hspace*{\fill}\\[10pt]
$\mQ_{jn}^{\lv_2 (3)}$	&	$\sum_{\mu,q \neq j} \frac{i \lambda_{qn}^\mu}{2} \mP_{\mu,q}^X \left( \Omega_{\mu,q}^* \ev{\sigma_{g_\mu e}^q} + \; \text{c.c.} \right)$ \hspace*{\fill}
\end{tabular}
\caption{\label{tab:ions_partial_sums} Additional partial sums, over the ions rather than modes, to simplify notation and speed up computation.} 
\end{table}
%\vspace*{-5pt}

\begin{widetext}

\begin{eqnarray}
&&\frac{d}{dt}\ev{b_n \sigma_{g_1g_1}^j} = -i\omega_n \ev{b_n \sigma_{g_1g_1}^j}
-\frac{i}{2}\left( \Omega_{1,j}^* \ev{b_n \sigma_{g_1e}^j} - \Omega_{1,j} \ev{b_n \sigma_{eg_1}^j}  \right)
+ \Gamma_1 \left( \ev{b_n} - \ev{b_n \sigma_{g_1g_1}^j} - \ev{b_n \sigma_{g_2g_2}^j} \right)
-\frac{\Omega_{1,j}^*\lambda_{j,n}^1}{2}\ev{\sigma_{g_1e}^j} \nonumber\\
&& -\frac{1}{2} \left\{ \Omega_{1,j}^* \left( (\mP_{1,jn}^{d X})^*\ev{\sigma_{g_1e}^j}
+ P_{1,j}^X \ev{b_n \sigma_{g_1e}^j} 
+ P_{1,j}^{X \sigma_{g_1e}}\ev{b_n} 
- 2 P_{1,j}^X \ev{b_n} \ev{\sigma_{g_1e}^j} \right) 
+ \Omega_{1,j} \left( \vphantom{P_{1,j}^{X \sigma_{g_1e}}} g_1e \rightarrow eg_1 \right) \right\}
+ \mQ_{jn}^{\lv_1} \ev{\sigma_{g_1g_1}^j} \nonumber\\
&&+ \frac{i\Omega_{1,j}^*\lambda_{jn}^1}{2} \mP_{1,j}^{X \sigma_{g_1e}}
+ \frac{i}{4} \left\{ \Omega_{1,j}^* \left( \mP_{1,j}^{X X} \ev{b_n \sigma_{g_1e}^j}
+ 2 (\mP_{1,jn}^{d X})^* \mP_{1,j}^{X \sigma_{g_1e}}
- 2 (\mP_{1,j}^X)^2 \ev{b_n} \ev{\sigma_{g_1e}^j} \right) 
- \Omega_{1,j} \left( \vphantom{\mP_{1,j}^{X \sigma_{g_1e}}} g_1e \rightarrow eg_1 \right)\right\}\nonumber\\
&& + \mQ_{jn}^{\lv_2 (1),\sigma_{g_1g_1}} 
+ \left( \mQ_{jn}^{\lv_2 (2)} - \mQ_{jn}^{\lv_2 (3)} \right) \ev{\sigma_{g_1g_1}^j}
\end{eqnarray}

\begin{eqnarray}
&&\frac{d}{dt}\ev{b_n \sigma_{g_1g_2}^j} = i\left( \Delta_{1,j}(t)-\Delta_{2,j}(t)-\omega_n\right)\ev{b_n \sigma_{g_1g_2}^j} 
+ \frac{i\Omega_{1,j}}{2}\ev{b_n \sigma_{eg_2}^j} 
- \frac{i\Omega_{2,j}^*}{2}\ev{b_n \sigma_{g_1e}^j}
-\frac{\Omega_{2,j}^*\lambda_{jn}^2}{2}\ev{\sigma_{g_1e}^j} \nonumber\\
&&-\frac{\Omega_{1,j}}{2} \left( (\mP_{1,jn}^{d X})^*\ev{\sigma_{g_2e}^j}^*
+ P_{1,j}^X \ev{b_n \sigma_{eg_2}^j} 
+ P_{1,j}^{X \sigma_{eg_2}}\ev{b_n} 
- 2 P_{1,j}^X \ev{b_n} \ev{\sigma_{g_2e}^j}^*   \right) \nonumber\\
&&-\frac{\Omega_{2,j}^*}{2} \left( (\mP_{2,jn}^{d X})^*\ev{\sigma_{g_1e}^j}
+ P_{2,j}^X \ev{b_n \sigma_{g_1e}^j} 
+ P_{2,j}^{X \sigma_{g_1e}}\ev{b_n} 
- 2 P_{2,j}^X \ev{b_n} \ev{\sigma_{g_1e}^j}   \right)
+ \mQ_{jn}^{\lv_1} \ev{\sigma_{g_1g_2}^j} \nonumber\\
&&+ \frac{i\Omega_{2,j}^*\lambda_{jn}^2}{2}\mP_{2,j}^{X \sigma_{g_1e}}
-\frac{i\Omega_{1,j}}{4} \left(  \mP_{1,j}^{X X} \ev{b_n \sigma_{eg_2}^j}
+ 2 (\mP_{1,jn}^{d X})^* \mP_{1,j}^{X \sigma_{eg_2}}
- 2 (\mP_{1,j}^X)^2 \ev{b_n} \ev{\sigma_{g_2e}^j}^* \right) \nonumber\\
&&+\frac{i\Omega_{2,j}}{4} \left(  \mP_{2,j}^{X X} \ev{b_n \sigma_{g_1e}^j}
+ 2 (\mP_{2,jn}^{d X})^* \mP_{2,j}^{X \sigma_{g_1e}}
- 2 (\mP_{2,j}^X)^2 \ev{b_n} \ev{\sigma_{g_1e}^j} \right) 
+ \mQ_{jn}^{\lv_2 (1),\sigma_{g_1g_2}} 
+ \left( \mQ_{jn}^{\lv_2 (2)} - \mQ_{jn}^{\lv_2 (3)} \right) \ev{\sigma_{g_1g_2}^j}
\end{eqnarray}

\begin{eqnarray}
&&\frac{d}{dt}\ev{b_n \sigma_{g_1e}^j} = -\left( \frac{\Gamma}{2}-i\left(\Delta_{1,j}(t)-\omega_n\right) \right)\ev{b_n \sigma_{g_1e}^j} 
-\frac{i\Omega_{1,j}}{2} \left( 2 \ev{b_n \sigma_{g_1g_1}^j} + \ev{b_n \sigma_{g_2g_2}^j} - \ev{b_n}  \right) -\frac{i\Omega_{2,j}}{2}\ev{b_n \sigma_{g_1g_2}^j} \nonumber\\
&&
+ \frac{\Omega_{1,j}}{2} \left\{ 2\left( (\mP_{1,jn}^{d X})^*\ev{\sigma_{g_1g_1}^j}^*
+ P_{1,j}^X \ev{b_n \sigma_{g_1g_1}^j} 
+ P_{1,j}^{X \sigma_{g_1g_1}}\ev{b_n} 
- 2 P_{1,j}^X \ev{b_n} \ev{\sigma_{g_1g_1}^j}^* \right) \right. 
 \left. + \left( \vphantom{P_{1,j}^{X \sigma_{g_1g_1}}} g_1g_1 \rightarrow g_2g_2 \right) 
- (\mP_{1,jn}^{d X})^* \right\} \nonumber\\
&& +\frac{\Omega_{1,j}\lambda_{jn}^1}{2}\ev{\sigma_{g_1g_1}^j}
+\frac{\Omega_{2,j}\lambda_{jn}^2}{2}\ev{\sigma_{g_1g_2}^j} 
+ \frac{\Omega_{2,j}}{2} \left( (\mP_{2,jn}^{d X})^*\ev{\sigma_{g_1g_2}^j}
+ P_{2,j}^X \ev{b_n \sigma_{g_1g_2}^j} 
+ P_{2,j}^{X \sigma_{g_1g_2}}\ev{b_n} 
- 2 P_{2,j}^X \ev{b_n} \ev{\sigma_{g_1g_2}^j}   \right)
+ \mQ_{jn}^{\lv_1} \ev{\sigma_{g_1e}^j} \nonumber\\
&&+ \frac{i\Omega_{1,j}}{4} \left\{  2 \left( \mP_{1,j}^{X X} \ev{b_n \sigma_{g_1g_1}^j}
+ 2 (\mP_{1,jn}^{d X})^* \mP_{1,j}^{X \sigma_{g_1g_1}}
- 2 (\mP_{1,j}^X)^2 \ev{b_n} \ev{\sigma_{g_1g_1}^j} \right)
+ \left( \vphantom{\mP_{1,j}^{X \sigma_{g_1g_1}}} g_1g_1 \rightarrow g_2g_2 \right) \right. 
\nonumber\\
&&\hphantom{+ \frac{i\Omega_{1,j}}{4} \{} \left.
- \left( \vphantom{\mP_{1,j}^{X \sigma_{g_1g_1}}} \mP_{1,j}^{X X}\ev{b_n} + 2 \mP_{1,jn}^{d X})^* \mP_{1,j}^X - 2 (\mP_{1,j}^X)^2 \ev{b_n}    \right) \right\}
+\frac{i\Omega_{1,j}\lambda_{jn}^1}{2} \mP_{1,j}^{X \sigma_{g_1g_1}}
+\frac{i\Omega_{2,j}\lambda_{jn}^2}{2} \mP_{2,j}^{X \sigma_{g_1g_2}} \nonumber\\
&&+ \frac{i\Omega_{2,j}}{4} \left( \mP_{2,j}^{X X} \ev{b_n \sigma_{g_1g_2}^j}
+ 2 (\mP_{2,jn}^{d X})^* \mP_{2,j}^{X \sigma_{g_1g_2}}
- 2 (\mP_{2,j}^X)^2 \ev{b_n} \ev{\sigma_{g_1g_2}^j}\right)
+ \mQ_{jn}^{\lv_2 (1),\sigma_{g_1e}} 
+ \left( \mQ_{jn}^{\lv_2 (2)} - \mQ_{jn}^{\lv_2 (3)} \right) \ev{\sigma_{g_1e}^j}
\end{eqnarray}

\begin{eqnarray}
&&\frac{d}{dt}\ev{b_n \sigma_{eg_1}^j} = -\left( \frac{\Gamma}{2}+i\left(\Delta_{1,j}(t)+\omega_n\right) \right)\ev{b_n \sigma_{eg_1}^j} 
+\frac{i\Omega_{1,j}^*}{2} \left( 2 \ev{b_n \sigma_{g_1g_1}^j} + \ev{b_n \sigma_{g_2g_2}^j} - \ev{b_n}  \right) 
+\frac{i\Omega_{2,j}^*}{2}\ev{b_n \sigma_{g_2g_1}^j} \nonumber\\
&&+ \frac{\Omega_{1,j}^*}{2} \left\{ 2\left( (\mP_{1,jn}^{d X})^*\ev{\sigma_{g_1g_1}^j}^*
+ P_{1,j}^X \ev{b_n \sigma_{g_1g_1}^j} 
+ P_{1,j}^{X \sigma_{g_1g_1}}\ev{b_n} 
- 2 P_{1,j}^X \ev{b_n} \ev{\sigma_{g_1g_1}^j}^* \right) \right. 
 \left. + \left( \vphantom{P_{1,j}^{X \sigma_{g_1g_1}}} g_1g_1 \rightarrow g_2g_2 \right) 
- (\mP_{1,jn}^{d X})^* \right\} \nonumber\\
&&+\frac{\Omega_{1,j}^*\lambda_{jn}^1}{2}\left( \ev{\sigma_{g_1g_1}^j} + \ev{\sigma_{g_2g_2}^j} - 1 \right) 
+ \frac{\Omega_{2,j}^*}{2} \left( (\mP_{2,jn}^{d X})^*\ev{\sigma_{g_1g_2}^j}^*
+ P_{2,j}^X \ev{b_n \sigma_{g_2g_1}^j} 
+ P_{2,j}^{X \sigma_{g_2g_1}}\ev{b_n} 
- 2 P_{2,j}^X \ev{b_n} \ev{\sigma_{g_1g_2}^j}^* \right) 
+ \mQ_{jn}^{\lv_1} \ev{\sigma_{g_1e}^j}^*\nonumber\\
&&- \frac{i\Omega_{1,j}}{4} \left\{  2 \left( \mP_{1,j}^{X X} \ev{b_n \sigma_{g_1g_1}^j}
+ 2 (\mP_{1,jn}^{d X})^* \mP_{1,j}^{X \sigma_{g_1g_1}}
- 2 (\mP_{1,j}^X)^2 \ev{b_n} \ev{\sigma_{g_1g_1}^j} \right)
+ \left( \vphantom{\mP_{1,j}^{X \sigma_{g_1g_1}}} g_1g_1 \rightarrow g_2g_2 \right) \right. 
\nonumber\\
&&\hphantom{- \frac{i\Omega_{1,j}^*}{4} \{} \left.
- \left( \vphantom{\mP_{1,j}^{X \sigma_{g_1g_1}}} \mP_{1,j}^{X X}\ev{b_n} + 2 \mP_{1,jn}^{d X})^* \mP_{1,j}^X - 2 (\mP_{1,j}^X)^2 \ev{b_n} \right) \right\}
-\frac{i\Omega_{1,j}^*\lambda_{jn}^1}{2}\left( \mP_{1,j}^{X \sigma_{g_1g_1}} + \mP_{1,j}^{X \sigma_{g_2g_2}} - \mP_{1,j}^X \right) \nonumber\\
&& - \frac{i\Omega_{2,j}^*}{4} \left( \mP_{2,j}^{X X} \ev{b_n \sigma_{g_2g_1}^j}
+ 2 (\mP_{2,jn}^{d X})^* \mP_{2,j}^{X \sigma_{g_2g_1}}
- 2 (\mP_{2,j}^X)^2 \ev{b_n} \ev{\sigma_{g_1g_2}^j}^*\right)
+ \mQ_{jn}^{\lv_2 (1),\sigma_{eg_1}} 
+ \left( \mQ_{jn}^{\lv_2 (2)} - \mQ_{jn}^{\lv_2 (3)} \right) \ev{\sigma_{g_1e}^j}^*
\end{eqnarray}

\end{widetext}

\bibliography{aps_template_eit_cooling_theory_paper}

\end{document}